\documentclass{lmcs}
\pdfoutput=1

\usepackage{lastpage}
\lmcsdoi{17}{3}{27}
\lmcsheading{}{\pageref{LastPage}}{}{}%
{Feb.~12,~2020}{Sep.~20,~2021}{}

\usepackage[utf8]{inputenc}

\usepackage{tikz,subcaption}
\usetikzlibrary{arrows,automata,shapes,positioning}
\usepackage{xstring,xspace}

\usepackage{MnSymbol}
\usepackage{multirow}
\usepackage{amsmath,amsthm,amsfonts}
\usepackage{comment}

\newenvironment{Property}[1]{\textbf{#1.} }{}

\newcommand{\match}{SELECT~}

\newcommand{\N}{{\mathbb{N}}}

\newcommand{\ECRPQ}{ECRPQ\xspace}
\newcommand{\ECRPQs}{ECRPQs\xspace}
\newcommand{\ECRPQLC}{ECRPQ+LC\xspace}

\newcommand{\pr}{PR}
\newcommand{\pra}{PRA}
\newcommand{\opra}{O\pra{}}
\newcommand{\piszczalka}[3]{#1 \to^{#2} #3}
\newcommand{\piszczalkaN}[4]{#1 \to^{#2}_{#4} #3}
\newcommand{\bnfpiszczalkaN}[4]{#1 {\color{terminal}\to}^{#2}_{#4} #3}

\definecolor{code}{RGB}{230,230,230}
\definecolor{terminal}{RGB}{200,30,30}
\newcommand{\bnfterm}[1]{{\color{terminal}\textbf{#1}}}
\newcommand{\bnfmath}[1]{{\color{terminal}\ensuremath{\mathbf{#1}}}}

\newcommand{\q}[1]{\colorbox{code}{\color{black}{\small{\texttt{#1}}}}}
\newcommand{\qbox}[1]{\noindent\colorbox{code}{\parbox{0.98\textwidth}{{{\color{black}{\small{\texttt{#1}}}}}}
    }\\
}

\renewcommand{\nc}[1]{\langle#1\rangle}
\newcommand{\labelling}[1]{\mathrm{#1}}
\newcommand{\comb}[1]{W(#1)} 
\newcommand{\qam}{QAM}
\newcommand{\naw}[1]{\langle#1\rangle}
\newcommand{\aut}{\mathcal{A}}
\newcommand{\padding}{\square}

\newcommand{\fmin}{\textsc{Min}}
\newcommand{\fmax}{\textsc{Max}}
\newcommand{\fcount}{\textsc{Count}}
\newcommand{\fsum}{\textsc{Sum}}

\newcommand{\op}{\triangle}
\newcommand{\definedBy}[2]{\mathrel{{#1}{:=}{#2}}}
\newcommand{\lang}{\mathcal{L}}
\newcommand{\Tvalue}[1]{\eta^G(#1)} 

\newcommand{\Z}{\mathbb{Z}}
\newcommand{\F}{{\mathcal{F}}}
\newcommand{\E}{{\mathrm{E}}}
\newcommand{\Erev}{{\mathrm{E}}^{-1}}
\newcommand{\lOne}{{\mathrm{One}}}
\newcommand{\M}{M}
\newcommand{\z}[2]{\IfEqCase{#2}{{-1}{\mathbf{Prev}(#1)}{+0}{#1}{+1}{\mathbf{Next}(#1)}}[\PackageError{z}{Undefined option to z: #2}{}]} 

\newcommand{\set}[1]{\{#1\}}

\newcommand{\tuple}[1]{{\langle#1\rangle}}
\newcommand{\Paragraph}[1]{\subparagraph{{\bf #1}}}

\newcommand{\NP}{\textsc{NP}}
\newcommand{\NL}{\textsc{NL}}

\newcommand{\PSpace}{\textsc{PSpace}}

\makeatletter
\newcommand\incircbin
{\mathpalette\@incircbin
}
\newcommand\@incircbin[2]
{\mathbin {\ooalign{\hidewidth$#1#2$\hidewidth\crcr$#1\bigcirc$}}}
\newcommand{\otop}{\incircbin{\top}}

\makeatother

\newcommand{\depth}{\textsc{depth}}

\newcommand{\pwi}[1]{\textcolor{green}{#1}}

\newcommand{\result}[2]{#1[#2]}

\newcommand{\reg}[1]{\mathbf{#1}}
\newcommand{\emb}[1]{#1^{\mathbf{se}}}
\newcommand{\demb}[1]{#1^{\mathbf{sed}}}
\newcommand{\A}{\mathcal{A}}

\usepackage{booktabs}

\begin{document}

\title{Modular Path Queries with Arithmetic}
\titlecomment{\lsuper*The work was supported by Polish National Science Centre grant 2014/15/D/ST6/00719.}
\author{Jakub Michaliszyn}
\author{Jan Otop}
\author{Piotr Wieczorek\texorpdfstring{\vspace{-5mm}}{}}
\address{University of Wroclaw, Poland}
\email{\{jmi,jotop,piotrek\}@cs.uni.wroc.pl}


\begin{abstract}
We propose a new approach to querying graph databases.
Our approach balances competing goals of expressive power,
language clarity and computational complexity.
A distinctive feature of our approach is the ability to express properties of minimal (e.g.\ shortest) and maximal (e.g.\ most valuable) paths satisfying given criteria.
To express complex properties in a modular way, we introduce  labelling-generating ontologies.
The resulting formalism is computationally attractive --- queries can be answered in non-deterministic logarithmic space in the size of the database.
\end{abstract}

\maketitle

\section{Introduction}\label{sec:intro}
Graphs are one of the most natural representations of data in a number of important applications such as modelling transport networks, social networks, technological networks (see the surveys~\cite{Angles+17,Wood12,Baeza13}). The main strength of graph representations is the possibility to naturally represent the connections among data.
Effective search and analysis of graphs is an important factor in reasoning performed in various tasks.
This motivates the study of query formalisms, which are capable of expressing properties of paths.

Nevertheless, most real-world data still resides in relational databases and relational engines are still the most popular database management systems~\cite{db-engines18}. Hence, it would be desirable to consider a query formalism that directly generalizes the relational approach, offers a natural representation of data values and, at the same time, enables convenient querying of the graph structure.

Modern day databases are often too large to be stored in computers' memory.
To make a query formalism computationally feasible, its query evaluation problem should be, preferably, in logarithmic space w.r.t.\ the size of the database (data complexity)~\cite{Calvanese+06a,Artale+07a,Barcelo+12}.
Checking existence of a path between two given nodes is already \NL{}-complete and hence \NL{} is the best possible lower bound for an expressive language for graph querying.
It is worth to mention that every problem in \NL{} can be solved deterministically in $O(\log^2 (n))$ space, which is a reasonable bound even for huge databases.

\begin{figure}
\centering
\begin{tikzpicture}[->,>=stealth',shorten >=1pt,auto,node distance=2cm,semithick]
  \tikzstyle{every state}=[fill=white,draw=black,text=black,minimum size=0.9cm]

  \node[state, ellipse] (E) at (-1,0)        {ECRPQ};
  \node[state, ellipse] (ELC) at (3,0)                   {ECRPQ+LC};
  \node[state, ellipse] (PR) at (2,2) {PR};
  \node[state, ellipse] (PRA) at (5, 2) {\pra};
  \node[state, ellipse] (OPRA) at (8, 2) {\opra{}};

  \path (E)  edge              (ELC)
             edge   	      node {$\subsetneq$} (PR)
        (PR) edge 	          node {$\subsetneq$} (PRA)
	    (PRA)edge 		      node {$\subsetneq$} (OPRA)
        (ELC)edge             node {$\subsetneq$} (PRA)
             edge             node[above right] {$\not \subseteq$, $\not \supseteq$} (PR)
;
\end{tikzpicture}
\caption{Comparison between different query languages. \ECRPQ is strictly subsumed by \pr{} by Theorem~\ref{th:subsumesECRPQ}, \ECRPQLC is strictly subsumed by \pra{} by Theorem~\ref{th:subsumesECRPQLC}, \pr{} and \ECRPQLC are incomparable by Theorem~\ref{th:incomparable}, \opra{} strictly subsumes \pra{} by Remark~\ref{rem:OpraStronger}, and \pra{} strictly subsumes \pr{} because it contains \ECRPQLC{} whilst \pr{} does not.
}%
\label{fig:diagram}
\end{figure}
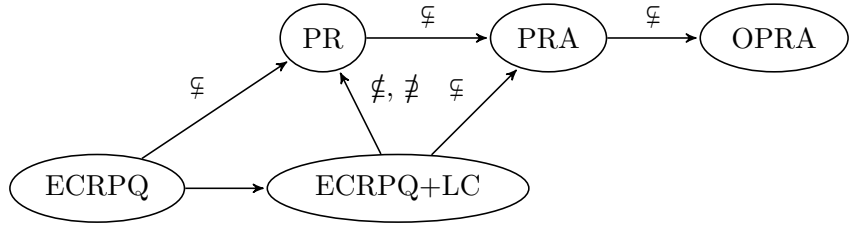

\Paragraph{Our contribution}
We propose a new approach for querying graph data, in which labelling-generating ontologies are first-class citizens.
It can be integrated with many existing query formalisms, both relational and graphical. To make the presentation clear we introduce the concept by defining a new language \opra{}. \opra{} features \NL{}-data complexity, good expressive power and a modular structure.
The expressive power of \opra{} strictly subsumes the expressive power of popular existing formalisms with the same complexity (see Fig.~\ref{fig:diagram}). Distinctive properties expressible in \opra{} are based on aggregation of data values along paths and computation of extremal values among aggregated data. One example of an \opra{}-expressible property is ``$p$ is a path from $s$ to $t$ that has both the minimal weight and the minimal length among all paths from $s$ to $t$''.
Within our model, our \opra{} queries can also use labelling functions defined as classic relational views by SQL queries.

Our data model is fairly general. The database consists of a finite number of nodes and a number of labelling functions assigning integers to fixed-size vectors of nodes. We assume for simplicity that the data values are integers; if needed, the labelling functions can be generalized to other datatypes.
Relational tables (of any arity) can be viewed as labelling functions assigning the value $1$ to the vectors in the given relation and $0$ otherwise.
In particular, classic edges can be represented implicitly by a binary labelling function returning $1$ for all pairs of nodes connected by an edge and $0$ otherwise, and weighted edges can be represented by allowing other values.
Nodes labelled by integers can be represented using unary labellings.

We define \opra{} in Section~\ref{sec:opra} in stages starting with two fragments of \opra{}, which we also employ later on in the discussion on the expressive power.
The first fragment is \pr{}, whose main components are the two types of constraints: \emph{Path} and \emph{Regular}.
We use path constraints to specify endpoints of graph paths; the other constraints only specify properties of paths. Regular constraints specify paths using regular expressions, adapted to deal with multiple paths and infinite alphabets.

The second fragment is \pra{}, that extends \pr{} with the \emph{Arithmetical} constraints.
The arithmetical constraints compare linear combinations of aggregated values, i.e., values of \emph{labels} accumulated along whole paths.
The language \pra{} can only aggregate and compare the values of labelling functions already defined in the graph. The properties we are interested in often require performing some arithmetical operations on the labellings, either simple (taking a linear combination) or complicated (taking minimum, maximum, or even computing some subquery).
Such operations are often nested inside regular expressions (as in~\cite{Grabon+16}) making queries unnecessarily complicated. Instead, similarly as in~\cite{Arenas+2014}, we specify such operations in a modular way as \emph{ontologies}. This leads to the language \opra{}, which comprises \emph{Ontologies} and \pra{}.
In our approach all knowledge on graph nodes, including all data values, is encoded by labellings. Our ontologies are also defined as \emph{auxiliary} labellings.
For example, having a labelling $\textrm{child}(x, y)$ stating that $x$ is a child of $y$, we can define a labelling $\textrm{descendant}(x, y)$ stating that $x$ is a descendant of $y$.
Such labellings can be computed on-the-fly during the query evaluation.

Then, in Section~\ref{sec:examples}, we present a number of examples intended as an illustration of the versatility and usefulness of the language. We also discuss the closure properties of \opra{} and possible applications of \opra{} to the verification of properties of  Kripke structures.

In Section~\ref{sec:ep}, we compare the expressive power of our formalisms
and the classic query languages
\emph{extended Conjunctive Regular Path Queries} (ECRPQ) and ECRPQ with \emph{linear constraints} (ECRPQ+LC)~\cite{Barcelo+12}.
We prove the main results depicted in Figure~\ref{fig:diagram}, primarily that \pr{} subsumes ECRPQ and \pra{} subsumes ECRPQ+LC\@.
We also illustrate there the additional expressive power of \opra{} over \pra{}.

Finally, in Section~\ref{sec:summing}, we study the complexity of the query evaluation problem for \opra{}. Namely, when the depth of nesting of auxiliary labellings is bounded, the problem is \PSpace{}-complete and its data complexity is \NL{}-complete.

This work is based on two conference papers~\cite{Grabon+16} and~\cite{Michaliszyn+17}.
The goal of~\cite{Grabon+16} is to design a query language for graphs that is able to express various arithmetic and aggregative
properties of nodes and paths, assumed that nodes are labelled with natural numbers. The resulting language can express properties like ``there is a path from $u$ to $v$ with the sum of nodes' values less than $c$''; however, it only works because the labels are non-negative. The language also cannot compare averages or even compare values of different paths.
The paper~\cite{Grabon+16} introduces one of the most important technical tools we use here: Query Applying Machines, which are Turing machines designed to provide a succinct representation of graphs that allow us to operate on them in logarithmic space.
The second paper~\cite{Michaliszyn+17} successfully removed many of these limitations, while keeping the good (\NL{}) data complexity.
In~\cite{Michaliszyn+17}, we introduced the language \opra{}, which we also use in this paper.
The language \opra{} operates on integer labels and can compare values of different paths, which allows us to study properties of best/longest/shortest paths.
In this paper, we extend the contribution of~\cite{Grabon+16} and~\cite{Michaliszyn+17} in a few ways. First of all, the paper~\cite{Michaliszyn+17} was written for graph databases, where we deal with a database that consists of a single graph.
Here, we adjust the narrative to present \opra{} as graph language that can be used with graph databases but also with relational databases, as by design it can easily deal with relations of arity greater than $2$.
We made the presentation of the language easier and provided some additional examples. We also added a study on the closure properties of \opra{} and a careful study on the expressive power on the language (there was a short discussion on the expressive power of \opra{} in~\cite{Michaliszyn+17}, but it focused mostly on the fact that \opra{} has access to more properties of the graph; e.g., \opra{} is more expressive than ECRPQ+LC because ECRPQ+LC cannot access nodes' labels; here we provide an analysis that does not depend on such technicalities).
Finally, we sharpen the complexity a bit: in~\cite{Michaliszyn+17}, we proved that for a bounded number of auxiliary labellings the data complexity is \NL{}-complete and the combined complexity is \PSpace{}-complete, whereas here we show that the data complexity is always \NL{}-complete, and the combined complexity is \PSpace{}-complete for queries with bounded nesting depth. We also provide here the full proofs of the complexity bounds.

\Paragraph{Design choices}
We briefly discuss the design choices we made during the development of \opra{}. Our main goal was to extend the expressive power of graph queries to express properties exemplified in Section~\ref{sec:examples}.
Once we overcame the main technical difficulties, the natural question arose: should we incorporate this expressive power into some existing language, either  practical or academic, or create a new language?
Our initial decision was to do the latter, and so we designed the language presented in the paper~\cite{Grabon+16}, which can be seen as a non-conservative extension of \ECRPQ{} (i.e., an extension increasing the expressive power).
This language, however, gained some negative feedback because of the complexity of the language, that obscured the main features we presented.
So here, having the lesson learned, we designed a new, modular language that, while is based on some ideas from \ECRPQ, is much simpler and therefore provides much easier insight into the expressive power we provide.
Due to the modular composition of the language, it is possible to integrate its features with other languages; for example, one can use \opra{} on top of a relational database that normally uses SQL\@.
It is also possible to integrate the distinctive features of \opra{} directly into the other languages, if needed.

Another choice we made in this paper was to abstract from the technical details of the implementation of \opra{} and to focus on the computational complexity aspect. There is an ongoing work on an implementation of \opra{} where these details are to be addressed.

\Paragraph{Related work} \emph{Regular Path Queries (RPQs)}~\cite{sigmod/CruzMW87,kr/CalvaneseGLV00} are usually used as a basic construction for graph querying.
RPQs are of the form $\piszczalka{x}{\pi}{y} \wedge \pi \in L(e)$ where $e$ is a (standard) regular expression.
Such queries return pairs of nodes $(v, v')$ connected by a path $\pi$ such that the labelling of $\pi$ forms a word from $L(e)$. \emph{Conjunctive Regular Path Queries (CRPQs)} are the closure of RPQs under conjunction and existential quantification~\cite{pods/ConsensM90,siamcomp/MendelzonW95}.
Barcelo et al.,~\cite{Barcelo+12} introduced \emph{extended CRPQs (ECRPQs)} that can compare tuples of paths by \emph{regular relations}~\cite{Elgot1965,tcs/FrougnyS93}. Examples of such relations are path equality, length comparisons, prefix (i.e., a path is a prefix of another path) and fixed edit distance. Regular relations on tuples of paths can be defined by the standard regular expressions over the alphabet of tuples of edge symbols.

Graph nodes often store \emph{data values} from an infinite alphabet. In such graphs, paths are interleaved sequences of data values and edge labels.
This is closely related to \emph{data words} studied in XML context~\cite{NevenSV04,DemriLN07,Segoufin06,BojanczykDMSS11}.
  Data complexity of query evaluation for most of the formalisms for data words is NP-hard~\cite{Libkin+16}.
This is not the case for \emph{register automata}~\cite{tcs/KaminskiF94}, which inspired Libkin and Vrgo\v{c} to define \emph{Regular Queries with Memory (RQMs)}~\cite{Libkin+16}. RQMs are again of the form $\piszczalka{x}{\pi}{y} \wedge \pi \in L(e)$, where $e$ is a \emph{Regular Expression with Memory (REM)}. REMs can store in a register the data value at the current position and test its equality with other values already stored in registers.
Register Logic~\cite{BarceloFL15} is, essentially, the language of REMs closed under Boolean combinations, node, path and register-assignment quantification. It allows for comparing data values in different paths. The positive fragment of Register Logic, RL$^+$, has data complexity in \NL{}, even when REMs can be nested using a branching operator.

Another related formalism is
Walk Logic~\cite{HellingsKBZ13} (WL), which extends FO with path quantification and equality tests of data values on paths. The main disadvantage of the Walk Logic is high complexity: query evaluation for WL is decidable but its data complexity is not elementary~\cite{BarceloFL15}.

\Paragraph{Aggregation}
Ability to use aggregate functions such as sum, average or count is a~fundamental mechanism in database systems.
Klug~\cite{Klug82} extended the relational algebra and calculus with aggregate functions and proved their equivalence.
Early graph query languages $G^+$~\cite{CruzMW88} or GraphLog~\cite{pods/ConsensM90,ConsensM93} can aggregate data values.
Consens and Mendelzon~\cite{ConsensM93} studied  \emph{path summarization}, i.e., summarizing information along paths in graphs.
They assumed natural numbers in their data model and allowed to aggregate summarization results.
In order to achieve good complexity (in the class NC) they allowed aggregate and summing operators that form a closed semiring. Other examples of aggregation can be found in~\cite{Wood12}.

Summing vectors of numbers along graph paths have been already studied in the context of various formalisms based on automata or regular expressions
and lead to a number of proposals that have combined complexity in PSPACE and data complexity in \NL{}.
Kopczy\'nski and To~\cite{KopczynskiT10} have shown that \emph{Parikh images} (i.e., vectors of letter counts) for the usual finite automata can be expressed
using unions of linear sets that are polynomial in the size of the automaton and exponential in the alphabet size (the alphabet size, in our context, corresponds to the dimension of vectors).
Barcelo et al.~\cite{Barcelo+12} extended ECRPQs with linear constraints on the numbers of edge labels counts along paths.
They expressed the constraints using reversal-bounded counter machines, translated further to Presburger arithmetic formulas of a polynomial size and evaluate them using techniques from~\cite{KopczynskiT10,Scarpellini84}.

Figueira and Libkin~\cite{FigueiraLibkin15} studied \emph{Parikh automata} introduced in~\cite{KlaedtkeR03}. These are finite automata that additionally store a vector of \emph{counters} in $\mathbb{N}^k$. Each transition also specifies a vector of natural numbers.
While moving along graph paths according to a transition the automaton adds this transition's vector to the vector of counters.
The automaton accepts if the computed vector of counters is in a given semilinear set in $\mathbb{N}^k$.
Also, a variant of regular expressions capturing the power of these automata is shown.
This model has been used to define a family of variants of CRPQs that can compare tuples of paths using \emph{synchronization languages}~\cite{FigueiraL15}.
This is a relaxation of regularity condition for relations on paths of ECRPQs and
leads to more expressive formalisms with data complexity still in \NL{}.
These formalisms are incomparable to ours since they can express non-regular relations on paths like suffix but cannot express properties of data values, nodes' degrees or extrema.

Cypher~\cite{TheNeo4jTeam18} is a practical query language first implemented in the graph database engine Neo4j.
It uses \emph{property graphs} as its data model. These are graphs with labelled nodes and edges, but edges and nodes can also store attribute values for a set of \emph{properties}.
\texttt{MATCH} clause of Cypher queries allows for specifying graph patterns that depend on nodes' and edges' labels as well as on their properties values.
OpenCypher, an initiative to standardize the language, has produced Cypher Query Language Reference (Version 9)~\cite{openCypher17}. More on Cypher can be found in the surveys~\cite{Angles+17} and~\cite{Francis+18a}.

G-Core~\cite{Angles+18} is a joint effort of industrial and academic partners to define a language that is composable (i.e.\ graphs are inputs and outputs of a query), treats paths as first-class citizens and integrates the most important features of existing graph query languages. The G-Core data model extends property graphs with paths. Namely, in a graph, there is also a (possibly empty) collection of paths. The paths have their identity and can have their own labels and $\langle$property, value$\rangle$ pairs. G-Core includes also features like aggregation and (basic) arithmetic along paths that is closely related with our proposal.  G-Core allows for defining costs of paths either by the hop-count (length) or by the positive weights (which may be computed by functional expressions). The full cost of a path is the sum of the weights and G-Core is able to look for paths that minimize it. In contrast, our data model allows negative weights.

In Section~\ref{sec:comparison} we compare the language constructions of G-Core, Cypher and \opra{} in more detail.

Another proposal of a graph query language of commercial strength is PGQL~\cite{Rest+16}. PGQL closely follows syntactic structures of SQL and defines powerful regular expressions that allow for filtering nodes and edges along paths as well as computing shortest paths.

Our data model allows to operate on property graphs~\cite{Angles+17}, where many edges between a pair of nodes are allowed.
To do so, for each of the edges we introduce a single, unique, additional node.
Then, an edge can be represented by a binary labelling returning $1$ for the pair of the source node and the additional node, and for the pair of the additional node and the target node for the edge. Naturally, the property values for the edge are assigned by unary labellings (named after the property keys) of the additional node.
We present an example of how to encode a property graph in Section~\ref{sec:prelim}.
Alternatively, an edge can be represented by a ternary labelling function returning 1 for all triples of the source, the target node and the corresponding additional node.

RDF~\cite{Cyganiak+2014} is a W3C standard that allows encoding of the content on the Web in a form of a set of \emph{triples} representing an edge-labelled graph.
Each triple consists of the subject $s$, the predicate $p$, and the object $o$ that
are resource identifiers (URI's), and represents an edge from $s$ to $o$ labelled by $p$.
Interestingly, the middle element, $p$, may play the role of the first or the third element of another triple.
Our formalism \opra{} allows operating directly on RDF without any complex graph encoding, by using a ternary labelling representing RDF triples.
This allows for convenient navigation by regular expressions, in which also the middle element of a triple can serve as the source or the target of a single navigation step (cf.~\cite{Libkin+18}).
The standard query formalism for RDF is SPARQL~\cite{PrudhommeauxS08,HarrisS+13}. It implements \emph{property paths}, which are RPQs extended with inverses and limited form of negation (see the survey~\cite{Angles+17}).

Another idea in processing graphs, fundamentally different than our approach is based on Pregel model~\cite{Malewicz+10} where the computation is organized in rounds. Basically, each node has a state and in each round is able to send messages to its neighbouring nodes and change its state according to the messages sent by its neighbours.

 \section{Preliminaries}%
\label{sec:prelim}

\begin{figure}
\centering
\begin{minipage}{0.48\textwidth}
\begin{tikzpicture}[->,>=stealth',shorten >=1pt,auto,node distance=2.9cm,semithick,scale=0.8]
  \tikzstyle{every state}=[fill=white,draw=black,text=black,minimum size=0.8cm]
  \tikzset{
   datanode/.style = {
    draw,
    rectangle,
    rounded corners,
    align=left,
    text width=1.85cm,
    inner ysep = +0.8em},
    labelnode/.style = {
    draw,
    rectangle,
    rounded corners,
    align=center,
    fill=white}
  }

\foreach \lab\linestyle\lX/\lY/\name/\kind/\cost/\time/\attr in
         {A/solid/0/0/S/square/0/10/5,
          AB1/dashed/3/2/T/tram/5/10/40,
          B/solid/6/4/P/park/100/60/30,
          AB2/dashed/6/0/W/walk/0/100/10,
          BA/dashed/0/4/B/bus/10/15/-2}
{

  \node[datanode,\linestyle] (\lab) at (\lX,\lY)
   {
type:\kind\newline time:\time\newline attr:\attr \vspace{-5pt}
   };
  \node[labelnode,\linestyle,fill=white,text width=0.6cm] (small\lab) at (\lab.north)
  {
   \name
  };

}

\path (A) edge[-] (AB1)
      (AB1) edge (B)
      (A) edge[-,out=350,in=190] (AB2)
      (AB2) edge[bend right] (B)
      (B) edge[-,out=170,in=10] (BA)
      (BA) edge[bend right] (A)
;

\end{tikzpicture}
\caption{A property graph $G$: the properties of the edges are in dashed boxes.}\label{fig:propertygraph}
\end{minipage}\hspace{8pt}
\begin{minipage}{0.48\textwidth}
\centering
\begin{tikzpicture}[->,>=stealth',shorten >=1pt,auto,node distance=2.9cm,semithick,scale=0.8]
  \tikzstyle{every state}=[fill=white,draw=black,text=black,minimum size=0.8cm]
  \tikzset{
   datanode/.style = {
    draw,
    rectangle,
    rounded corners,
    align=left,
    text width=1.85cm,
    inner ysep = +0.8em},
    labelnode/.style = {
    draw,
    rectangle,
    rounded corners,
    align=center,
    fill=white}
  }

\foreach \lab\lX/\lY/\name/\kind/\cost/\time/\attr in
         {A/0/0/S/square/0/10/5,
          AB1/3/2/T/tram/5/10/40,
          B/6/4/P/park/100/60/30,
          AB2/6/0/W/walk/0/100/10,
          BA/0/4/B/bus/10/15/-2}
{

  \node[datanode] (\lab) at (\lX,\lY)
   {
type:\kind\newline time:\time\newline attr:\attr \vspace{-5pt}
   };
  \node[labelnode,fill=white,text width=0.6cm] (small\lab) at (\lab.north)
  {
   \name
  };

}

\path (A) edge (AB1)
      (AB1) edge (B)
      (A) edge[out=350,in=190] (AB2)
      (AB2) edge[bend right] (B)
      (B) edge[out=170,in=10] (BA)
      (BA) edge[bend right] (A)

;
\end{tikzpicture}
\caption{An equivalent graph $G'$: its nodes represent the nodes and the edges of $G$.}\label{fig:map}
\end{minipage}
\end{figure}

Various kinds of data representations for graphs are possible and presented in the literature. The differences typically include the way the elements of graphs are labelled --- both  nodes and edges may be labelled by finite or infinite alphabets, which may have some inner structure. Here, we choose a general approach in which a \emph{labelled graph}, or simply a graph, is a tuple consisting of a finite number of \emph{nodes} $V$ and a number of labelling functions $\lambda: V^l \to \Z\cup\{-\infty, \infty\}$ assigning integers to vectors of nodes of some fixed size (we allow labellings without parameters, i.e., $l=0$, for reasons that will be apparent later).

While edges are not explicitly mentioned, if needed, one can consider an \emph{edge} labelling $\E$ such that $\E(v, v')$ is $1$ if there is an edge from $v$ to $v'$ and it is $0$ otherwise.

A special case of a labelled graph is a \emph{relational database}. In this case, the range of all the labellings is $\{0, 1\}$ (and hence the labellings can be called ``relations''), and every node has to be in at least one relation. Clearly, each of the labellings defines a relation of the same arity as the labelling.

For convenience, we assume that the set of nodes contains a distinguished node $\padding$ --- this is an artificial node we use to avoid problems with paths of different lengths.
Note that all labellings must define their values also for tuples with $\padding$ and we do not restrict these values in principle.

A \emph{path} is a sequence of nodes. For a path $p=v_1 \dots v_k$, by  $|p|$ we denote the length of $p$ and by $p[i]$ we denote its $i$-th element, $v_i$, if $i\leq k$, and $\padding$ otherwise.

To compare our language with other formalisms, we will also consider a special subclass of labelled graphs, called \emph{standard graphs}. These are graphs with a single unary labelling function $\lambda: V \to \set{0, \dots, k}$ for some $k$ and a single binary labelling function $\E: V^2 \to \set{0, 1}$. We show how to represent \emph{property graphs}~\cite{Angles+17} in our model. In this model nodes and edges of a graph can be annotated with properties of the form of key-value pairs. We present an example of a property graph $G$ in Figure~\ref{fig:propertygraph}. The graph $G$ represents a fragment of a map. The nodes represent \emph{places} and the edges represent \emph{links} between the places. Both the nodes and the edges of $G$ contain labels such as $S$ or $T$ and a number of properties, e.g., a type which identifies what kind of a place, or a link, it is. As depicted in Figure~\ref{fig:map}, and already discussed in Section~\ref{sec:intro}, we represent both nodes and edges of property graphs as nodes in our model.
This is hardly a surprise since we adopt a lightweight concept of edges. The  binary labelling $\E$ we use, yields a value 0 or 1 for a pair of nodes specifying the existance of an edge and does not store any additional information. Note that the key-value pairs are also represented with labellings, in particular each labelling is named after the key.

\section{Language \opra{}}\label{sec:opra}
To define \opra{}, we first introduce its two fragments, \pr{} and \pra{}, which can be seen as syntactic restrictions of \opra{}.
The first fragment, \pr{}, includes \emph{Path constraints} and \emph{Regular constraints}, and the second one, \pra{}, includes also \emph{Arithmetical constraints}.

The language \opra{} will extend \pra{} with \emph{Ontologies}.
For simplicity, we assume that the queries return only node identifiers. For practical purposes this can be straightforwardly extended to returning labels.

We now introduce basic notational conventions we use.
By $x$ we denote a node variable, by $\pi$ a path variable, by $\vec{x}$ and $\vec{\pi}$ tuples of node and path variables respectively, by $c$ an integer value. By $\lambda$ we denote a labelling; for binary labellings (i.e.\ with two arguments), we often use symbol $E$  instead of $\lambda$. Our conventions are summarised in the following listing.

\qbox{
\begin{tabular}{ll}
$x, x_1, \ldots \in \mathtt{NodeVariables}$ & node variables  \\
$\pi, \pi_1, \ldots \in \mathtt{PathVariables}$ & path variables \\
$c, c_1, \ldots \in \mathtt{Integers}$ & integer values\\
$E, E_1, \ldots \in \mathtt{BinaryLabellings}$ & binary labellings\\
$\lambda, \lambda_1, \ldots \in \mathtt{Labellings}$ & labellings (of any arity)\\
$v, t, \ldots \in V$& nodes\\
$p, p_1, \ldots \in V^*$& paths\\
$f, g, f_1, \ldots  \in \F$& functions\end{tabular}
}

\opra{} queries are of the form (red font distinguishes the terminal symbols):

\qbox{
\noindent
\hspace{-1cm}
$
\left.
\begin{tabular}{l}
\:\quad
$Q(\vec{x}, \vec{\pi})$ ::= \bnfterm{LET} \text{O}ntologies \bnfterm{IN} \\
$\left.
\begin{tabular}{l}
$\left.
\begin{tabular}{l}
\bnfterm{\match NODES} $\vec{x}$\bnfterm{, PATHS} $\vec{\pi}$ \\
\bnfterm{SUCH THAT} \text{P}athConstraints \\
\bnfterm{WHERE} \text{R}egularConstraints
\end{tabular}
\quad \right] \: \pr{}$ \\
\:\! \bnfterm{HAVING} \text{A}rithmeticalConstraints
\end{tabular}
\quad \right] \: \pra{}$ \\
\end{tabular}
\quad \right] \: \opra{}
$
}
where $\vec{x}$ are free node variables, $\vec{\pi}$ are free path variables,
\q{PathConstraints} is a~conjunction of \emph{path constraints}, \q{RegularConstraints} is a~conjunction of \emph{regular constraints}, \q{ArithmeticalConstraints} is a~conjunction of \emph{arithmetical constraints}, and \q{\text{O}ntologies} is a sequence of \emph{auxiliary labellings definitions} defined in Section~\ref{lang:opra}. Either of the tuples $\vec{x}, \vec{\pi}$ can be empty.
The conjuncts in these constraints are connected with the keyword \q{AND}:

 \qbox{
\begin{tabular}{lll}
PathConstraints &::=& PathConstraint [ \bnfterm{AND} PathConstraints ] \\
RegularConstraints &::=& RegularConstraint [ \bnfterm{AND} RegularConstraints ]  \\
ArithmeticalConstraints &::=& ArithmeticalConstraint\\
& & [ \bnfterm{AND} ArithmeticalConstraints ]
\end{tabular}
}

The constraints may contain variables not listed in the \q{SELECT} clause (which are then existentially quantified).
Unnecessary components may be omitted (e.g., the keyword \q{WHERE} if no regular constraints are needed).
Each node variable may be also treated as a path variable, representing a single-node path.

\subsection{Path and regular constraints}
The language \pr{} is a syntactic fragment of \opra{} obtained by disallowing the keywords \q{LET \dots{} IN} and \q{HAVING}.
In other words, \pr{} queries consist of \emph{path constraints},
which involve node and path variables, and \emph{regular constraints}, which involve path variables only.
We complete its definition by defining the path constraints and regular constraints.

\Paragraph{Syntax}
 A path constraint is an expressions of the form ${\piszczalkaN{x_s}{\pi}{x_t}{\E}}$, where $x_s, x_t$ are node variables, $\pi$ is a path variable, and $\E$ is a binary labelling, i.e.,\\
 \qbox{
\begin{tabular}{lll}
PathConstraint &::=&  $\bnfpiszczalkaN{x_s}{\pi}{x_t}{E}$
\end{tabular}
}
where $x_s, x_t \in \mathtt{NodeVariables}$, $\pi \in \mathtt{PathVariables}$, $E \in \mathtt{BinaryLabellings}$.

Formal syntax of regular constraints is defined in the following way.

\qbox{
\begin{tabular}{lll}
RegularConstraint &::=& \bnfmath{\epsilon} | NodeConstraintConj | RegularConstraint$^{\bnfmath{*}}$\\
&& | RegularConstraint \bnfmath{+} RegularConstraint \\&& | RegularConstraint \bnfmath{\cdot} RegularConstraint  \\
NodeConstraintConj & ::= & NodeConstraint [ \bnfmath{\wedge} NodeConstraintConj ] \\
NodeConstraint &::=  & $\bnfmath{\nc{\top}}$ | $\bnfmath{\nc{\texttt{\color{black}NCValue NCOperator NCValue}}}$ \\
NCValue &::= & $c$ | $\lambda$\bnfterm{(}PathPositionVariable, \dots, PathPositionVariable\bnfterm{)} \\
PathPositionVariable &::= & $\bnfmath{\z{{\color{black}\pi}}{-1}}$ | $\bnfmath{\z{{\color{black}\pi}}{0}}$ | $\bnfmath{\z{{\color{black}\pi}}{1}}$\\
NCOperator & ::= & $\bnfmath{\leq}$ | $\bnfmath{\leq}$ | $\bnfmath{<}$ | $\bnfmath{\geq}$ | $\bnfmath{>}$ | $\bnfmath{\neq}$
\end{tabular}
}
where $c \in \mathtt{Integers}$, $\pi \in \mathtt{PathVariables}, \lambda \in \mathtt{Labelligs}$.

Each node constraint has a natural number parameter $k$. Such $k$-node constraint is either $\nc{\top}$, denoting a dummy constraint that always holds or an expression of the form $\nc{X \op X'}$, where
$\op \in \set{\leq, <, =, >, \geq, \neq}$ and each of $X, X'$ is an integer constant or
 a labelling function $\lambda$ applied to some of the variables in $\Pi$.

 A regular constraint $R(\pi_1, \ldots, \pi_k)$ is syntactically a regular expression over an infinite alphabet consisting of conjunctions of all $k$-node constraints.
 We write regular expressions with $\epsilon$ denoting the empty word, $\cdot$ denoting \emph{concatenation}, $+$ denoting \emph{alternation}, and $^*$ denoting \emph{Kleene star}.

 Regular constraints are  evaluated over paths from \match part and paths quantified existentially.
In order to allow accessing nodes on a path $\pi$, we introduce fresh, free node variables $\z{\pi}{-1}, \z{\pi}{+0}$, and $\z{\pi}{+1}$.
Naturally, $\z{\pi}{-1}$, $\z{\pi}{+0}$ and $\z{\pi}{+1}$ represent the nodes at the previous, the current and the next position of $\pi$ accordingly.
By $\Pi$ we denote the set of the variables $\z{\pi}{-1}$, $\z{\pi}{+0}$, $\z{\pi}{+1}$ for all paths $\pi$ of a given query.

\Paragraph{Intuitions for Regular Constraints}
Before we define the semantics of regular constraints we present an intuition first. An important part of a query language for graphs is the way to specify graph patterns.
In \pr{}, we define them using conjunctions of regular constraints.
Regular constraints extend the formalism of  Regular Path Queries (RPQs) to deal with the values of labelling functions.
The idea is different than the one involving Regular Expressions with Memory (REMs)~\cite{Libkin+16}.
REMs can store the values at the current position in registers and then test their equality with other values already in registers.
Here, we limit \pr{} expressions to access  only the nodes at the current, the next and the previous position on each of the paths. Later on, in the last example of Section~\ref{sec:examples}, we show how \opra{} enables to compare values of nodes at any distant positions.

\Paragraph{Path Constraints Semantics}
Given a set of node and path variables $\mathcal{V}$ and a graph $G$ with the nodes $V$ we define a variable instantiation $\eta^G$ as a function from $\mathcal{V}$ to the nodes and paths of $G$.
We sometimes omit the superscript when the graph is clear from the context and we do not distiguish an instantiation
 and its canonical extention to tuples of node and path variables.

 Given a labelling $\lambda$ in a graph $G$ and a tuple $\vec{v}$ of nodes of $G$ (of appropriate arity)
  we say that $G \models \lambda(\vec{v})$ iff $\lambda(\vec{v}) \neq 0$.
   A path constraint $\piszczalkaN{x_s}{\pi}{x_t}{\E}$ \emph{holds} in a graph $G$ under an instatiation $\eta^G$, which we denote by  $G, \eta^G \models \piszczalkaN{x_s}{\pi}{x_t}{\E}$,
iff $\eta^G(\pi)$ is a sequence of nodes $v_0v_1\dots v_n$ of $G$ starting from $v_0=\eta^G(x_s)$, ending in $v_n=\eta^G(x_t)$
and such that for all $i \in \set{0,\ldots, n-1}, G \models \E(\eta^G(x_i), \eta^G(x_{i+1}))$.

\Paragraph{Regular Constraints Semantics}
Given paths $p_1, \dots, p_k$ we define $\comb{p_1, \dots, p_k} \in (V^{3k})^*$
to be the word of the length of the longest path among $p_1, \dots, p_k$ such that for each $j$ we have $\comb{p_1, \dots, p_k}[j] = (p_1[j-1], p_1[j], p_1[j+1], \ldots, p_k[j-1], p_k[j], p_k[j+1])$ (i.e., it is a vector consisting of three consecutive nodes of each path, substituted by $\padding$ if not present). For example, if $p_1 = v_1v_2v_3$ and $p_2=v_4v_5$ then $\comb{p_1, p_2} = ((\padding, v_1, v_2, \padding, v_4, v_5), (v_1, v_2, v_3, v_4, v_5, \padding)$,
 $(v_2, v_3, \padding, v_5, \padding, \padding))$.

Intuitively, the semantics is given by applying the specified labelling functions
to the nodes in the given vector and comparing the values according to the $\op$ symbol.
A node constraint may be seen as a function that takes a vector of $3k$ nodes (i.e., $\comb{p_1, \dots, p_k}[j]$, if $j$ is the current position on the paths), represented by the variables in $\Pi$, and returns a Boolean value.

For an example consider a conjunction $c$ of two node constraints
\[\nc{\lambda(\z{\pi_1}{+0}) > \lambda(\z{\pi_1}{+1}) \wedge \lambda(\z{\pi_1}{-1}) = \lambda(\z{\pi_2}{+1})},\]
let $i$ be the current position and for some $\eta$ let \[\comb{\eta(\pi_1), \eta(\pi_2)}[i] = ((v^1_{i-1}, v^1_i, v^1_{i+1}), (v^2_{i-1}, v^2_i, v^2_{i+1})).\]
Then $c$ requires that $v^1_i>v^1_{i+1}$ and $v^1_{i-1}=v^2_{i+1}$.

Given a graph $G$ over vertices $V$, the language of $R$, denoted as $L^G(R)$, is a subset of ${(V^{3k})}^*$ defined using the usual rules: $L^G(\epsilon)$ is the empty word, for a conjunction of $k$-node constraints $r$, $L^G(r)$ is the set of all vectors of length $3k$ for which $r$ returns true, and $L^G(R \cdot R')$, $L^G(R + R')$ and $L^G(R^*)$ are defined inductively as the concatenation, the union and Kleene star closure of appropriate languages.

Given a graph $G$ and an instatiation $\eta^G$ we say that $R(\pi_1, \dots, \pi_k$) holds in $G$ under $\eta^G$, $G, \eta^G \models R(\pi_1, \dots, \pi_k$), if and only if $\comb{\eta^G(\pi_1), \dots, \eta^G(\pi_k)} \in L^G(R)$.

\Paragraph{Example} Consider the query:

\qbox{\match NODES $x, y$\\
SUCH THAT $\piszczalkaN{x}{\pi}{y}{\E}$\\
WHERE $\nc{\E(\z{\pi}{+1}, \z{\pi}{+0})}^*\nc{\top}$
	AND $\nc{\lambda(\z{\pi}{+0}) > 0}^*$
}

Let $G$ be a graph and let $\eta^G$ be an instatiation. Let $v = \eta^G(x)$ and $w = \eta^G(y)$.
The query holds in $G$ iff there is a bidirectional path between $v$ and $w$ whose each node is labelled by $\lambda$ with a positive number. Notice that $\nc{\top}$ is required as $\z{\pi}{+1}$ is $\padding$ for the last node.
If $G$ has an additional binary labelling $\Erev$ such that for all nodes $v, w$ we have $\Erev(v, w) = \E(w, v)$, then the same query can be stated as follows:\\
\qbox{\match NODES $x, y$\\
SUCH THAT $\piszczalkaN{x}{\pi}{y}{\E}$ AND $\piszczalkaN{x}{\pi}{y}{\Erev}$\\
WHERE $\nc{\lambda(\z{\pi}{+0}) > 0}^*$
}
Note that the existantially quantified path $\pi$ is mentioned in both of the path constraints above.
 We discuss in Section~\ref{lang:opra} how to define $\Erev$ as an auxiliary labelling based on $\E$.

\Paragraph{Query semantics}
 Let $Q(\vec{x}, \vec{\pi})$ be a \pr{} query. Given a graph $G$ and tuples of nodes $\vec{v}$ and paths $\vec{p}$ of $G$ we say that $Q(\vec{v}, \vec{p})$ holds in $G$, denoted by $G \models Q(\vec{v}, \vec{p})$, if and only if there exists an instantiation $\eta^G$ of all free and existential node and path variables of $Q$ such that $\eta^G(\vec{x}) = \vec{v}$, $\eta^G(\vec{\pi}) = \vec{p}$ and
all path and regular constraints of $Q$ hold for $G$ under $\eta^G$.

\subsection{Arithmetical constraints}%
\label{lang:pra}
The syntax for arithmetical constraints is defined as follows.

\qbox{
\begin{tabular}{lll}
ArithmeticalConstraint &::=& ArithmeticalSum \bnfmath{\leq} $d$ \\
ArithmeticalSum &::=& $c$ ArithmeticalAtom [ \bnfmath{+} ArithmeticalSum ]\\
ArithmeticalAtom &::=& $\lambda$ \bnfterm{\bf [} $\pi_1\bnfterm{,} \ldots\bnfterm{,} \pi_n$ \bnfterm{\bf ]}
\end{tabular}
}
where $c \in \mathtt{Integers}$, $d  \in \mathtt{Integers}$ or $d$ is an  parameterless labelling, $\pi_1, \ldots, \pi_n \in \mathtt{PathVariables}$ and  $\lambda \in \mathtt{Labellings}$.

The language \pra{} is a syntactic fragment of \opra{} obtained by disallowing the keyword \q{LET \dots{} IN}, i.e., it extends \pr{} with arithmetical constraints, which we define below.

An \emph{arithmetical atom} is of the form $\lambda[\pi_{1}, \ldots, \pi_{k}]$, where $\lambda$ is a labelling. An \emph{arithmetical constraint} is an inequality
\( c_{1} \Lambda_1 + \cdots + c_j \Lambda_j\leq d \),
where $c_1, \ldots, c_j$ are integer constants, $d$ is either an integer constant or a  parameterless labelling, and each $\Lambda_\ell$ is an arithmetical atom.

\Paragraph{Semantics}
Let $\Lambda$ be an arithmetical atom $\lambda[\pi_{1}, \ldots, \pi_{k}]$ and let $G$ be a graph.
Consider an instantiation $\eta^G$ of path variables $\pi_{1}, \ldots, \pi_{k}$ with paths $p_1, \dots, p_k$. The value of $\Lambda$ under $\eta^G$, denoted by $\mathrm{val}_{\eta^G}(\Lambda)$, is defined as
\[
 \sum_{i=1}^{s} \lambda(p_1[i], \dots, p_k[i])
\]
where $s=\max(|p_1|, \dots, |p_k|)$.

The arithmetical constraint $c_{1} \Lambda_1 + \cdots + c_j \Lambda_j \leq d$ holds in $G$ under $\eta^G$ iff
the value $c_{1} \mathrm{val}_{\eta^G}(\Lambda_1) + \cdots + c_j \mathrm{val}_{\eta^G}(\Lambda_j)$ is less than or equal to $d$.

\Paragraph{Example}
Consider a class of graphs with an edge labelling $\E$, a unary labelling $\lOne$, which returns $1$ for all nodes, and a unary labelling $\lambda$.
The following query holds in a graph $G$ under an instatiation $\eta^G$ when the nodes $\eta^G(x), \eta^G(y)$ are connected by a path $p = \eta^G(\pi)$ such that each node of $p$ is labelled by $\lambda$ with a positive number and the average value of $\lambda$ over all nodes is
at most $5$.

\qbox{\match NODES $x, y$\\
SUCH THAT $\piszczalkaN{x}{\pi}{y}{\E}$\\
WHERE $\nc{\lambda(\z{\pi}{+0}) > 0}^*$\\
HAVING $\lambda[\pi] - 5\cdot {\lOne}[\pi] \leq 0$
}

\Paragraph{Query semantics}
The query semantics is defined in a virtually the same way as for \pr{} queries; we additionally require the instantiation to satisfy all arithmetical constraints of the query.

\subsection{Auxiliary labellings}%
\label{lang:opra}
The language \opra{} extends \pra{} with the constructions that define auxiliary labellings of graphs, which are defined based on existing graph labellings and its structure.

\qbox{
\begin{tabular}{lll}
Ontologies &::=& LabellingDef [ \bnfterm{,} Ontologies  ] \\
LabellingDef &::=& $\lambda$\bnfterm{(}$\vec{x}$\bnfterm{)} \bnfmath{:=} LabellingDefTerm\bnfterm{(}$\vec{x}$\bnfterm{)} \\
LabellingDefTerm($\vec{x}$) & ::= & $c$  | $\lambda$\bnfterm{(}$\vec{y}$\bnfterm{)} |
$x_1$ \bnfterm{=} $x_2$ | \bnfterm{\bf [}$Q$\bnfterm{(}$\vec{x}$\bnfterm{)\bf ]} | \\
& &  $f$\bnfterm{(}LabellingDefTerm\bnfterm{(}$\vec{y}_1$\bnfterm{)}, $\ldots$, LabellingDefTerm\bnfterm{(}$\vec{y}_k$\bnfterm{)}\bnfterm{)}  | \\
& &  $\min_{\lambda, \pi} Q$\bnfterm{(}$\vec{x}, \pi$\bnfterm{)} |
 $\max_{\lambda, \pi} Q$\bnfterm{(}$\vec{x}, \pi$\bnfterm{)} |\\
& & $f$\bnfterm{(\{} LabellingDefTerm$\bnfterm{(}x\bnfterm{)}$ \bnfmath{\colon} LabellingDefTerm$\bnfterm{(}x\bnfterm{,} \vec{y}\bnfterm{)}$\bnfterm{\})}
\end{tabular}
}
where $Q$ is an \opra{} query, $c$ is an integer, $-\infty$ or $+\infty$, $\pi \in \mathtt{PathVariables}$, $x$ is a fresh node variable (not in $\vec{x}$), $\vec{x}$ is a tuple of node variables, $x_1, x_2 $ are variables from $\vec{x}$ and tuples $\vec{y}, \vec{y}_1, \ldots, \vec{y}_k$ consist of the variables from $\vec{x}$, $\lambda \in \mathtt{Labellings}$.

The ability to define auxiliary labellings significantly extends the expressive power of the language.
The essential property of auxiliary labellings is that their values do not need to be stored in the database, which would require polynomial memory, but can be computed \emph{on demand}. An auxiliary labelling may be seen as an \emph{ontology} or a \emph{view}.

\q{Ontologies} is of the form  $\definedBy{\lambda_1(\vec{x}_1)}{t_1}, \ldots, \definedBy{\lambda_n(\vec{x}_n)}{t_n}$.
Such a sequence defines auxiliary labellings $\lambda_1, \dots, \lambda_n$ that can be used in the \pra{} query and also in the following labellings, i.e., $\lambda_i$ can be used in $\lambda_j$ if $i<j$. The labellings are defined by means of \emph{terms} $t_1, \dots, t_n$, which are expressions with free variables $\vec{x}_1, \dots, \vec{x}_n$.

To define terms, we assume a set $\F = \F_\A \cup \F_b$ of \emph{fundamental functions}
\[f:{(\Z\cup\{-\infty, \infty\})}^* \to \Z\cup\{-\infty, \infty\},\] where
$\F_\A$ is a set of \emph{aggregate functions} including maximum $\fmax$, minimum $\fmin$,
counting $\fcount$, summation $\fsum$, and
$\F_b$ is a set of \emph{binary} functions including $+$, $-$, $\cdot$ and $\leq$.
The set $\F$ can be extended, if needed, by any functions computable by a non-deterministic Turing machine (see Remark~\ref{nondeterministicfuns}) whose size of all tapes while computing $f(\vec{x})$  is
logarithmic in length of $\vec{x}$ and values in $\vec{x}$, assuming binary representation, provided that additional aggregate functions in $\F$ are invariant under permutation of arguments.

We distinguish four types of terms.

\Paragraph{Basic terms} There are three basic types of terms: a constant $c$, a labelling value $\lambda(\vec{y})$, and a node identity test $y = y'$, where $y, y'$ are node variables, $\vec{y}$ is a vector of node variables, $c$ is a constant, $\lambda$ is a labelling.

\Paragraph{Function application}
A term can be a function application to another terms: $f(t_1, \dots, t_n)$, where $f \in \F$ and $t_1, \dots, t_n$ are terms.

\Paragraph{Subqueries}
There are two essential constructs involving subqueries in terms.
First, we can evaluate the truth value of a subquery, i.e., the expression $[Q(\vec{y})]$, where $Q$ is an \opra{} query with $\vec{y}$ being free node variables.
Note that $\vec{y}$ occur in the main quary and hence they become instantiated in the subquery. Therefore, $[Q(\vec{y})]$ returns the value $1$ if under the current instantiation of $\vec{y}$, query $Q(\vec{y})$ holds, and $0$ otherwise.
Second, we minimise (resp., maximise) the value of a parameter satisfying a subquery, i.e., the value of the expression $\min_{\lambda, \pi} Q(\vec{y}, \pi)$ (resp. $\max_{\lambda, \pi} Q(\vec{y}, \pi)$), where $Q$ is an \opra{} query with free node variables $\vec{y}$ and a single free path variable $\pi$, to obtain the minimum (resp.\ maximum) of values $\lambda[\pi]$ over paths satisfying $Q$ (as usual, the minimum of the empty set is $+\infty$ and maximum of the empty set is $-\infty$). The expression $\lambda[\pi]$ denotes as in the arithmetical constraints the value of the labelling $\lambda$ applied to the path $\pi$ (i.e., the sum of the values of $\lambda$ along $\pi$).

\Paragraph{Aggregative properties}
The last type of term allows us to apply a function to a set of labels of all nodes \emph{satisfying} some label.
The syntax is $f(\{ t(x) \colon t'(x,\vec{y})\})$, where $f\in \F_\A$ is an aggregate function.
The value can be defined as follows: first, we compute the set $X=\set{x_1, \dots, x_s}$ of nodes such that for every $x \in X$ we have $t'(x, \vec{y}) = 1$.
Then, the value of the term is the value of $f(t(x_1), t(x_2), \dots, t(x_s))$. Notice that since $f$ is an aggregate function, the value does not depend on the order among $x_1, \dots, x_s$.

\Paragraph{Formal semantics}
The semantics of terms is as follows. Let $G$ be a graph and let $\eta^G$ be an instatiation.
In Table~\ref{t:tvalue} we inductively extend instantiations to terms.
If $G$ is clear from the context, we write $t(\vec{v})$ as a shorthand of $\eta^G(t(\vec{x}))$, where $\eta^G(\vec{x}[i])=\vec{v}[i]$ for all $i$.

\begin{table}[!h]
\begin{tabular}{cp{110mm}}
Term $t$ & The value of $ \Tvalue{t}$\\
\toprule
$c$ & $c$, where $c$ is a constant \\[1mm]
${\lambda(\vec{y})}$ & $\lambda(\eta^G(\vec{y}))$, where $\lambda$ is a labelling of $G$\\
${[Q(\vec{y})]}$ & $1$ if $Q(\eta^G(\vec{y})
)$ holds in $G$ and $0$ otherwise,\\[1mm]
${\min_{\lambda, \pi} Q(\vec{y}, \pi)}$ &
the minimum of values of $\lambda[p]$, defined as in the arithmetical constraints, over all paths $p$ such that
$Q(\eta^G(\vec{y}),p)$ holds in $G$, \\[1mm]
${\max_{\lambda, \pi} Q(\vec{y}, \pi)}$ &
the maximum of values of $\lambda[p]$, defined as in the arithmetical constraints, over all paths $p$ such that
$Q(\eta^G(\vec{y}),p)$ holds in $G$, \\[1mm]
${y=y'}$ & $1$ if $\Tvalue{y}=\Tvalue{y'}$ and $0$ otherwise (a node identity check),\\[1mm]
${f(t_1(\vec{y}_1), \ldots, t_k(\vec{y}_k))} \) & the value of \( f(\Tvalue{t_1(\vec{y}_1)},\ldots, \Tvalue{t_k(\vec{y}_k}))$,  \\[1mm]
\({f(\{ t(x) {:} t'(x,\vec{y}) \})}\) & the value of \(f(t(v_1), \ldots, t(v_n))\),
where $v_1, \ldots, v_n$ are all nodes $v$ of $G$ satisfying $t'(v, \eta^G(\vec{y})) = 1$. \end{tabular}
\caption{Value of terms.}\label{t:tvalue}
\end{table}

\Paragraph{Auxiliary labellings}
Consider a term $t(\vec{x})$ of an arity $k$ and a graph $G$, which does not have a labelling $\lambda$. We define the graph $G[\definedBy{\lambda}{t}]$ as the graph $G$ extended with the labelling $\lambda$ of an arity $k$
such that
$\lambda(\vec{v}) = t(\vec{v})$ for all $\vec{v}\in V^k$.
We call $\lambda$ an \emph{auxiliary labelling} of $G$.
We write  $G[\definedBy{\lambda_1}{t_1}, \ldots, \definedBy{\lambda_n}{t_n}]$ to denote
the results of successively adding labellings $\lambda_1, \dots, \lambda_{n}$ to the graph $G$, i.e., $((G[\definedBy{\lambda_1}{t_1}])[\definedBy{\lambda_2}{t_2}])\ldots [\definedBy{\lambda_n}{t_n}]$.

\Paragraph{Size of an \opra{} query}
The size of a query $Q$ of the form \q{LET O IN Q'} is the sum of binary representations of terms $t_1, \ldots, t_n$ in \q{O}
and the size of the query $Q'$.

\Paragraph{Example}
We can define labellings $\Erev$ and $\lOne$ presented in the above examples with terms \q{$\Erev(x,y) := \E(y,x)$} and \q{$\lOne(x) := 1$}.
The example from Section~\ref{lang:pra} can be stated for graphs without $\lOne$ labelling as follows:

\qbox{LET $\lOne(x) := 1$ IN \\
\match NODES $x, y$\\
SUCH THAT $\piszczalkaN{x}{\pi}{y}{\E}$\\
WHERE $\nc{\lambda(\z{\pi}{+0}) > 0}$\\
HAVING $\lambda{}[\pi] - 5\cdot {\lOne}[\pi] \leq 0$
}

\Paragraph{Semantics of \opra{}} Let $Q(\vec{v}, \vec{p})$ be of the form $\q{LET O IN Q'}$.
Given a graph $G$ and tuples of nodes $\vec{v}$ and paths $\vec{p}$ of $G$ we say that $Q(\vec{v}, \vec{p})$ holds in $G$, denoted by $G \models Q(\vec{v}, \vec{p})$ if and only if $G[O] \models Q'(\vec{v}, \vec{p})$
Note that $Q'(\vec{x},\vec{\pi})$ is a \pra{} query, which can refer to auxiliary labellings $\lambda_{1}, \ldots, \lambda_{n}$.

\section{Examples}%
\label{sec:examples}

We focus on the following scenario: a graph database that corresponds to a map of some area.
Each graph's node represents either a \emph{place} or a \emph{link} from one place to another.
The graph has four unary labellings and one binary labelling.
The labelling  $\labelling{type}$ represents the type of a place for places (e.g., square, park, pharmacy) or the mode of transport for links (e.g., walk, tram, train); we assume each type is represented by a constant, e.g., $c_{\textrm{square}}$, $c_{\textrm{park}}$.
 The labelling $\labelling{attr}$ represents attractiveness  (which may be negative, e.g., in unsafe areas),
and $\labelling{time}$ represents time.
The binary labelling $\labelling{E}$ represents edges: for nodes $v_1, v_2$, the value $\labelling{E}(v_1,v_2)$ is $1$ if there is an edge from $v_1$ to $v_2$
and $0$ otherwise.
For example, the graph on Fig.~\ref{fig:map} represents a map with two places: $S$ is a square and $P$ is a park. There are three nodes representing links: node $W$ represents moving from $S$ to $P$ by walking, $T$ moving from $S$ to $P$ by tram and $B$ moving from $P$ to $S$ by bus.

\subsection{Language \pra}%
\label{subsec:pra}

\begin{Property}{Sums}
The language \pra{} can express properties of paths' sums.
Consider the query $Q_1(x, y)$ below. For nodes $s, t$, the query $Q_1(s, t)$ holds iff there is a route from $s$ to $t$ that takes at most 6 hours and its attractiveness is over 100.
\noindent \qbox{$Q_1(x, y)$ = \match NODES $(x, y)$
SUCH THAT $\piszczalkaN{x}{\pi}{y}{\labelling{E}}$
HAVING $\labelling{time}[\pi] \leq 360 \wedge \labelling{attr}[\pi] > 100$}

Furthermore, we can compute averages, to some extent. For example, the following arithmetical constraint says that for some path $\pi$ the average attractiveness of $\pi$ is at least 4 attractiveness points per minute:
 \q{$\labelling{attr}[\pi] \geq 4 \labelling{time}[\pi]$}.
\end{Property}

\begin{Property}{Multiple paths}
We define a query $Q_2(x, y)$ that asks whether there is a route from $x$ to $y$, such that from every place we can take a tram (e.g., if it starts to rain).
We express that by stipulating a route $\pi$ from $x$ to $y$ and a sequence $\rho$ of tram links, such that every node of $\pi$
representing a place is connected with the corresponding tram link in $\rho$. In a way, $\rho$ works as an existential quantifier for nodes of $\pi$.

\noindent \qbox{$Q_2(x, y)$ = \match NODES $(x, y)$
SUCH THAT $\piszczalkaN{x}{\pi}{y}{\labelling{E}}$
WHERE  $\nc{\labelling{type}(\z{\rho}{+0}) = c_{\mathrm{tram}}}^\ast \wedge \mathit{link}(\pi, \rho)$}
where the parameterized macro \q{$\mathit{link}(\pi, \rho)$} is defined as
\[{(\langle \labelling{type}(\z{\pi}{+0}) = c_{\mathrm{bus}} \rangle + \langle \labelling{type}(\z{\pi}{+0}) = c_{\mathrm{walk}} \rangle + \langle \labelling{type}(\z{\pi}{+0}) = c_{\mathrm{tram}} \rangle +\langle \labelling{E}(\z{\pi}{+0}, \z{\rho}{+0}) = 1 \rangle )}^\ast{}\] states that every node of the first path
either is not a place, i.e, it represents any of possible links (by a bus, a walk or a tram), or is connected with the corresponding node of the second path.
\end{Property}

\subsection{Language  \opra}%
\label{s:opra}
We show how to employ auxiliary labellings in our queries.
For readability, we introduce some syntactic sugar --- constructions which do not change the expressive power of \opra, but allow queries to be expressed more clearly.
We use the function symbols $=, \neq$ and Boolean connectives, which can be derived from $\leq$ and arithmetical operations. Also,
we use terms $t(x,y)$ in arithmetical constraints, which can be expressed by first defining the labelling $\definedBy{\labelling{\lambda}(x, y)}{t(x, y)}$,
defining additional paths $\rho_1 = x,\rho_2 = y$ of length $1$, and using $\labelling{\lambda}[\rho_1, \rho_2]$.

\begin{Property}{Processed labellings}
Online route planners often allow looking for routes which do not require much walking.
The query $Q_3(x, y)$ asks whether there exists a route from $x$ to $y$ such that the total walking time is at most 10 minutes.
To express it, we define a labelling $\labelling{t\_walk}(x)$, which is the time of $x$, if it is a walking link, and $0$ otherwise.

\noindent \qbox{$Q_3(x, y)$ = LET \(\labelling{t\_walk}(x) := (\labelling{type}(x) = c_\textrm{walk}) \cdot \labelling{time}(x)\) IN\\
\match NODES $(x, y)$ SUCH THAT $\piszczalkaN{x}{\pi}{y}{\labelling{E}}$ \\
HAVING \(\labelling{t\_walk}[\pi] \leq 10\)
}
\end{Property}

\begin{Property}{Nested queries}
It is often advisable to avoid crowded places, which are usually the most attractive places.
We write a query that holds for routes that are always at least 10 minutes away from any node with attractiveness greater than $100$.
We define a labelling  \q{$\labelling{crowded}(x)$} as

\noindent \qbox{$Q_4(x)$ =
 [\match NODES $(x)$ SUCH THAT  $\piszczalkaN{x}{\pi}{y}{\labelling{E}}$  \\
WHERE $\langle \top \rangle^\ast \langle \labelling{attr}(\z{\pi}{+0}) > 100 \rangle$
 HAVING  $\labelling{time}[\pi] \leq 10$]}

\noindent Notice that the variables $\pi$ and $y$ are existentially quantified. We check whether the value of \q{$\labelling{crowded}$} is $0$ for each node of the path $\pi$.

\noindent \qbox{$Q_5(\pi)$ = \match PATHS $(\pi)$
SUCH THAT $\piszczalkaN{x}{\pi}{y}{\labelling{E}}$
 WHERE $ \langle \labelling{crowded}(\z{\pi}{+0}) = 0\rangle^\ast$}
\end{Property}

\begin{Property}{Nodes' neighbourhood}
``Just follow the tourists'' is an advice given quite often. With \opra{}, we can verify whether it is a good advice in a given scenario.
A route is called \emph{greedy} if at every position, the following node on the path is the most attractive successor.
We define a labelling \q{$\labelling{MAS}(x,y)$} that is $1$ if $y$ is the most attractive successor of $x$, and $0$ otherwise,  and use it to express that there is a greedy route from a node $x$ to a node $y$.\\ \qbox{$Q_6(x, y)$ =
LET \( \labelling{MAS}(x, y) :=  (\fcount(\{  \labelling{attr}(z) \colon \labelling{E}(x, z) \land \labelling{attr}(z) {\geq} \labelling{attr}(y) \})  {=} 1)\)   IN\\
\match NODES \((x, y) \)
SUCH THAT \(\piszczalkaN{x}{\pi}{y}{\labelling{E}}\)\\
WHERE \( \langle \labelling{MAS}(\z{\pi}{+0}, \z{\pi}{+1}) = 1\rangle^\ast\langle\top\rangle \)
}

The above query considers routes that in each place select the most attractive link, and in each link select the most attractive place. What if we are interested in the attractiveness of the places only?
If we assume that $x$ and $y$ are places and all edges are between places and links only, then this can be achieved by replacing the \q{WHERE} clause of the above query by:
\qbox{WHERE \( {(\langle \top \rangle \langle  \labelling{MAS'}(\z{\pi}{-1}, \z{\pi}{+1}) = 1\rangle)}^\ast \langle\top\rangle \)
}
where \q{MAS'} is a counterpart of \q{MAS} obtained by replacing $E(x,z)$ by $E(x, x') \land E(x', z)$.
Such a query checks at every link that the following place is the most attractive place that can directly reached from the previous place.

\end{Property}

\begin{Property}{Properties of paths' lengths}
{In route planning, we can optimize different aspects such as time, necessary budget or attractiveness. We can express the conjunction of such objectives, i.e., specify routes that are optimal for several objectives.
The following query asks whether is it possible to get from $s$ to $t$ with a route that takes the shortest time among all routes, and at the same time it maximises the attractiveness among all routes.}

\noindent \qbox{$Q_8(x, y)$ = \match NODES $(x, y)$ SUCH THAT  $\piszczalkaN{x}{\pi}{y}{\labelling{E}}$\\
HAVING
$(\labelling{attr}[\pi] = \max_{\labelling{attr}, \rho} Q_\mathrm{route}(x, y, \rho)) \wedge
(\labelling{time}[\pi] = \min_{\labelling{time}, \rho} Q_\mathrm{route}(x, y, \rho))$}\\
where $Q_\mathrm{route}(x, y, \rho)$ stands for \q{\match NODES $(x, y)$, PATHS $\rho$ SUCH THAT $\piszczalkaN{x}{\rho}{y}{\labelling{E}}$}.
\end{Property}

\begin{Property}{Registers}
Registers are an important concept in graph query languages.
For instance, to express that two paths have a non-empty intersection, we load a (non-deterministically picked) node to a register and check whether it occurs in both paths.
This can be achieved by the following query.

\qbox{$Q_9(\pi_1, \pi_2)$ = LET $\labelling{loop}(x,y) := x=y$ IN \\
\match PATHS $\pi_1, \pi_2$\\
SUCH THAT $\piszczalkaN{z}{\pi}{z}{\labelling{loop}}$\\
WHERE $\mathit{same}(\pi_1, \pi) \land \mathit{same}(\pi_2, \pi)$
}
where $\mathit{same}(\rho_1, \rho_2) = \langle \top \rangle^*\langle \z{\rho_1}{+0} = \z{\rho_2}{+0} \wedge \z{\rho_1}{+0} \neq \padding \rangle \langle \top \rangle^*$. This query uses a constant register $\pi$.

The following query asks whether there exists a route from a club $s$ to a club $t$ on which the attractiveness of visited clubs never decreases.
In the register-based approach, we achieve this by storing the most recently visited club in a separate register.
Here, we express this register using an additional path $\rho$, storing the values of the register, a labelling
$\labelling{r}(x', y, y')$ which states that $y' = x'$ if $x'$ is a club, and $y'=y$ otherwise, and an auxiliary labelling $\otop$ which is true for all the pairs of nodes.

\qbox{$Q_{10}(x, y)$ = LET $\otop(x,y) := 1$,\\
\hspace{4em} $\labelling{r}(x', y, y') := {(\labelling{type}(x') = c_\textrm{club} \Rightarrow y' = x') \land (\labelling{type}(x') \neq c_\textrm{club} \Rightarrow y=y')}$
 IN \\
\match NODES $(x, y)$
SUCH THAT \( \piszczalkaN{x}{\pi}{y}{\labelling{E}}  \land
\piszczalkaN{x}{\rho}{y}{\labelling{{\otop}}}\) \\
WHERE
\(\mathit{ends}(\pi) \land \mathit{regs}(\pi, \rho) \land \mathit{inc}(\rho)\)}\\
where the macro \q{$\mathit{ends}(\pi) = \langle \labelling{type}(\z{\pi}{+0})=c_\textrm{club} \rangle \langle \top \rangle^\ast{} \langle \labelling{type}(\z{\pi}{+0}) = c_\textrm{club} \rangle$} states that both ends of a path are clubs,
\q{$\mathit{regs}(\pi, \rho)=
\langle \labelling{r}(\z{\pi}{+1},\z{\rho}{+0}, \z{\rho}{+1}) = 1 \rangle^\ast{}\langle \top \rangle$} ensures that at each position the second path contains the most recently visited club along the first path, and the part
\q{\( \mathit{inc}(\rho)=
 \langle\labelling{attr}(\z{\rho}{+0}) \leq \labelling{attr}(\z{\rho}{+1}) \rangle^\ast{}\langle \top \rangle
\)} checks that the attractiveness never decreases.
\fussy
\end{Property}

\newcommand{\store}[1]{[\hspace{-1pt}\xleftarrow{#1}\hspace{-1pt}]}

 \section{G-Core, Cypher and \opra}\label{sec:comparison}

In this section we compare constructs of  G-Core~\cite{Angles+18}, Cypher~\cite{TheNeo4jTeam18} and \opra.

All three formalisms have SQL-like clauses and rely crucially on matching of graph patterns.
Such patterns are specified in G-Core and Cypher using ASCII-art syntax, e.g., the pattern

\qbox{\noindent  (n)-[:connection]->(m)->n}

 binds the variables $n$, $m$ to the pairs of nodes $(a,  b)$ such that there is a directed edge labelled \verb+connection+ from $a$ to $b$ and there is also some edge going back. Patterns can be fixed structures (\emph{rigid} patterns), consisting of the exact graph that should match the input graph but also may be specified by navigational paths using regular expressions.
In \opra{} we provide two options: path constraints $\piszczalkaN{x_s}{\pi}{x_t}{E}$ that essentially define reachability over a binary labelling that encodes edges,
as well as powerful regular constraints.

The process of matching patterns generates tuples of matched node, edge and path values that can be filtered by WHERE conditions.
Technically, a graph-pattern match corresponds to a homomorphism from a query $Q$ to the input graph $G$.
All three formalisms allow for an unconstrained semantics where the multiple variables may match the same value (a node or an edge).
The default matching semantics in CYPHER is, however, \emph{no-repeated-edge semantics}~\cite{Angles+17} where variables corresponding to edges have to map one-to-one and the other variables need not to be mapped injectively.
 Each of the languages may produce an enumeration of all matched tuples as well as of some of the projections on subsets of variables.
G-Core and CYPHER support shortest paths (hop count) matching, G-Core allows also for weighted shortest paths where the costs may be specified using positive weights.

In CYPHER and G-Core the stream of tuples generated by pattern matching can be aggregated and then returned to be processed in the following parts of a query.
For example, consider the following CYPHER query~\cite{Angles+17} to find the longest movies in a collection.

\noindent \qbox{MATCH (m:Movie) WITH MAX(m.runtime) AS maxTime \\
MATCH (m:Movie) WHERE m.runtime = maxTime\\
RETURN m}
The first pattern \verb+MATCH (m:Movie)+ matches all movies, aggregates their lengths and return the maximal length using \verb+WITH+ clause. Then, the second \verb+MATCH (m:Movie)+
again matches all movies but this time, however, it filters out the ones with $\labelling{runtime}$ not equal to $\labelling{maxTime}$.

Although we do not allow in \opra{} for an aggregation at this stage we can reformulate such queries as follows.

\noindent  \qbox{LET \textrm{maxTime}() = $\fmax(\mathrm{runtime}(x) \mid \mathrm{type}(x) = c_{\mathrm{movie}})$ IN \\
\match NODES $(x)$  \\
WHERE $\nc{\mathrm{maxTime}() = \mathrm{runtime}(x) }$
}

Recall that here we use the fact that each node variable may be also treated as a path variable that represents a single-node path.

The specific thing for \opra{} is the ability to compare paths in terms of regular relations~\cite{Barcelo+12}.
Regular relations that can be specified in regular constrains include path equality, prefix (i.e., is a path a prefix of another?), length comparisons, fixed-edit distance, synchronous transformation.

G-Core and \opra{} have tractable data complexity, on the other hand there are at least two reasons for which data complexity of CYPHER is NP-hard.
We have already mentioned that CYPHER has no-repeated-edge semantics for graph pattern matching. This makes the evaluation intractable~\cite{siamcomp/MendelzonW95}.
 The second reason is its ability to unwind paths, that is to return path elements (e.g.\ nodes) and then to process them.
This feature may be used~\cite{Angles+17} to write a fixed query that returns two different disjoint routes between given two nodes which is also an NP-hard problem.
Note that this is precisely the reason for which we do not allow in \opra{} nested queries with free \emph{path} variables (only node variables are allowed).
We discuss this topic in Section~\ref{sec:closure}.

G-Core and CYPHER have a number of features that are not present in \opra. In particular, in \opra{} there are no features allowing for any modification of graphs nor their data values.
We can only define new labellings and then use them in the following part of queries. It is also not possible to construct and return graphs (and thus queries are not composable).

 \section{Closure properties}%
 \label{sec:closure}
In this section we discuss closure properties of \opra{} under standard set-theoretic operations.
We define these operations formally as follows.
First, for a query $Q(\vec{x},\vec{\pi})$ and a graph $G$ we define the \emph{result} of $Q$ on $G$, denoted by $\result{Q}{G}$, as
\[
\result{Q}{G} = \set{ (\vec{v},\vec{p}) \mid \vec{v} \in {(V\setminus \set{\padding})}^{|\vec{x}|}, \vec{p} \in {({(V\setminus \set{\padding})}^*)}^{|\vec{\pi}|}, G \models Q(\vec{v},\vec{p}) }
\]

To avoid problems with the artificial node $\padding$, which is used to align paths of different lengths, we ignore it in $\result{Q}{G}$.

\begin{itemize}
    \item a query $Q_\exists(x_{i_1}, \ldots x_{i_k}, \pi_{j_1}, \ldots, \pi_{j_l})$ is a \emph{projection} of $Q(\vec{x}, \vec{\pi})$ if and only if
    $i_1, \ldots, i_k$ are distinct indices from $\set{1, \ldots, |\vec{x}|}$, $j_1, \ldots, j_k$  are distinct indices from $\set{1, \ldots, \vec{|\pi|}}$, and
    for every graph $G$ we have $\result{Q_{\exists}}{G} = \set{ \tuple{v_{i_1}, \ldots, v_{i_k}, p_{j_1}, \ldots, p_{j_l}} \mid (\vec{v},\vec{p}) \in \result{Q}{G} }$,
    \item a query $Q_{\cap}(\vec{x},\vec{\pi})$ is an \emph{intersection} of $Q_1(\vec{x},\vec{\pi})$ and $Q_2(\vec{x},\vec{\pi})$ if and only if
    for every graph $G$ we have $\result{Q_{\cap}}{G} = \result{Q_1}{G} \cap \result{Q_2}{G}$,
    \item a query $Q_{\cup}(\vec{x},\vec{\pi})$ is a \emph{union} of $Q_1(\vec{x},\vec{\pi})$ and $Q_2(\vec{x},\vec{\pi})$ if and only if
    for every graph $G$ we have $\result{Q_{\cup}}{G} = \result{Q_1}{G} \cup \result{Q_2}{G}$,
    \item a query $Q_\times(\vec{x}, \vec{x}', \vec{p}, \vec{p}')$ is a \emph{Cartesian product} of $Q_1(\vec{x}, \vec{\pi}), Q_2(\vec{x}, \vec{\pi})$ if and only if
    $\vec{x}$ and $\vec{x}'$ are disjoint,  $\vec{\pi}$ and $\vec{\pi}'$ are disjoint, and
    for every graph $G$ we have \[\result{Q_\times}{G} = \set{ (\vec{v},\vec{v'}, \vec{p},\vec{p'}) \mid (\vec{v},\vec{p}) \in \result{Q_1}{G}, (\vec{v}',\vec{p}'') \in \result{Q_2}{G}}\]
    \item a query $Q_C(\vec{x},\vec{\pi})$ is a \emph{complement} of $Q$ if and only if
    \[\result{Q_C}{G} = {(V\setminus \set{\padding})}^{|\vec{x}|} \times {({(V\setminus \set{\padding})}^*)}^{|\vec{\pi}|} \setminus \result{Q}{G}.\]
\end{itemize}

\begin{thm}\label{t:closure}
Given \opra{} queries $Q_1, Q_2$, we can compute in polynomial time every projection of $Q_1$, the intersection, the union, and the Cartesian product of $Q_1$ and $Q_2$.
If $Q_1$ has no free path variables, then we can compute in polynomial time the complement of $Q_1$.
\end{thm}
\begin{proof}[Proof of Theorem~\ref{t:closure}]
The projection case is straightforward --- a projection of a query can be obtained by simply not listing the unwanted variables in the \q{\match} statement.

To define the complement we simply use $Q_1$ as a subquery (we can do that only for queries without free path variables).

\qbox{LET C($\vec{x}$) := [$Q_1(\vec{x})$] IN
\match NODES $\vec{x}$
HAVING C[$\vec{x}$]=0
}

Having queries $Q_1, Q_2$ with only node variables, we can give similar constructions for the cases of a Cartesian product, an intersection and a union.
For queries with free path variables, the constructions are more complex.

Assume that, for $i=1,2$ the query $Q_i$ is of the form

\qbox{LET $O_i$ IN\\
\match NODES $\vec{x}_i$, PATHS $\vec{\pi}_i$\\
SUCH THAT $P_i$\\
WHERE $R_i$\\
HAVING $A_i$}
Without loss of generality, we assume that quantified variables in $Q_1$ and $Q_2$ are disjoint (if not, we simply rename the conflicting entities).

For the Cartesian-product case, we assume that nodes variables $\vec{x}_1$ and $\vec{x}_2$ are disjoint, and path variables $\vec{\pi}_1$ and $\vec{\pi}_2$ are disjoint as well.
Then, the following query expresses $Q_{\times}$:

\qbox{LET $O_1$, $O_2$ IN\\
\match NODES $\vec{x}_1, \vec{x}_2$, PATHS $\vec{\pi}_1,\vec{\pi}_2$\\
SUCH THAT $P_1 $ AND $ P_2$\\
WHERE $R_1 $ AND $ R_2$\\
HAVING $A_1 $ AND $ A_2$}

For the intersection case, we assume that $\vec{x}_1 = \vec{x}_2$ and $\vec{\pi}_1  = \vec{\pi}_2$.
We construct the query $Q_{\cap}$ as follows:

\qbox{LET $O_1$, $O_2$ IN\\
\match NODES $\vec{x}_1$, PATHS $\vec{\pi}_1$\\
SUCH THAT $P_1 $ AND $ P_2$ \\
WHERE $R_1 $ AND $ R_2$\\
HAVING $A_1 $ AND $ A_2$}

Finally, the union case is the most difficult one. Roughly speaking, we want to do the Cartesian product of two queries and define the result as the union of the projections of this product. Assuming, without the loss of generality, that the variables in the queries are disjoint, this can be achieved in the following way.

\qbox{LET $O_1$, $O_2$ IN\\
\match NODES $\vec{x}$, PATHS $\vec{\pi}$\\
SUCH THAT $P_1 $ AND $ P_2$\\
WHERE $R_1$ AND $R_2$ AND $\mathrm{EQ}$\\
HAVING $A_1 $ AND $ A_2$}
where $\vec{x}$, $\vec{\pi}$ are fresh variables. The regular constraint  $\mathrm{EQ}$
guarantees that
either ($\vec{x}=\vec{x}_1$ and $\vec{\pi} = \vec{\pi}_1$), or ($\vec{x}=\vec{x}_2$ and $\vec{\pi} = \vec{\pi}_2$).
It can be defined in an \opra{} query in a straightforward way
using a regular constraint with alternation and a new auxiliary labelling defined with a node identity check; notice that the definition will depend on the arity of the vectors.

Notice however, if one of the queries is empty, the Cartesian product is empty, and this naive approach fails.

To avoid this problem, we first define, for each $i \in \set{1,2}$, an additional labelling $\lambda_{Q_i} = [Q_i]$. By definition, $\lambda_{Q_i}$ is 0 if $Q_i$ returns the empty result and $1$ otherwise.
Then, for each $i$, we define $R'_i$ based on $R_i$ in the following way: for each conjunct $r$ of $R_i$, $R_i'$ contains $r + \langle \lambda_{Q_i} = 0 \rangle^*$. Finally, for each $i$, we define $A'_i$ based on $A_i$ in the following way: for each conjunct $\sum_i s_i < d$ of $A_i$, $A'_i$ contains the conjunct $\sum_i s_i<\lambda_d$, where $\lambda_d$ is an auxiliary labelling equal to $d$ if $\lambda_{Q_i}=1$ and $\infty$ otherwise. This can be defined as $\min(d, (2\lambda_{Q_i}-1) \cdot \infty)$.

The above definition means that $R'_i$ and $A'_i$ are trivially satisfied if $\lambda_{Q_i}=0$. Let $O$ be the definitions of the auxiliary labellings described in the paragraphs above.
Putting it all together, we obtain:

\qbox{LET $O$, $O_1$, $O_2$ IN\\
\match NODES $\vec{x}$, PATHS $\vec{\pi}$\\
SUCH THAT $P_1 $ AND $ P_2$\\
WHERE $R'_1 $ AND $ R'_2 $ AND $\mathrm{EQ}$\\HAVING $A'_1 $ AND $ A'_2$}
\end{proof}

In Theorem~\ref{t:closure} the construction for the complement is given only for \opra{} queries without free path variables. We show that assuming that $\NL \neq \NP$, \opra{} queries are not closed under the complement.
Indeed, we show in the following Lemma~\ref{l:HamiltonianCycles} that if all \opra{} queries are closed under the complement, then
there exists a boolean query $Q_{\textrm{ham}}$ which holds if there is a Hamiltonian cycle in a graph.
However, we show in Theorem~\ref{th:query-in-NL} that for a fixed query, the query evaluation problem is $\NL$-complete.
Therefore, having the query $Q_{\textrm{ham}}$, we can decide the existence of a Hamiltonian cycle in a given $G$ in $\NL$ and hence $\NL = \NP$.

\begin{lem}%
\label{l:HamiltonianCycles}
Assume that all \opra{} queries are closed under the complement.
Then, there exists a boolean \opra{} query $Q_{\textrm{ham}}$ such that for every graph $G$,
the query $Q_{\textrm{ham}}$ holds in $G$ if and only if $G$ has a Hamiltonian cycle.
\end{lem}
\begin{proof}
The query $Q_{\textrm{ham}}$
is the conjunction of the following queries with a free path variable $\pi$ that becomes existentially quantified in $Q_{\textrm{ham}}$:
\begin{itemize}
\item
$Q_{\textrm{len}}(\pi)$ which holds for the cycles connected by $E$ with the length equal to the number of all nodes in a graph, and
\item
$Q_{\textrm{unique}}(\pi)$ which holds for the paths in which all nodes are different.
\end{itemize}
Note that such paths are Hamiltonian cycles.

The query $Q_{\textrm{len}}(\pi)$ is as follows

\qbox{LET $\lOne(x) := 1$, $\mathrm{Nodes}(x) := sum(\set{ \lOne(y) : \top })$ IN\\
\match PATHS $\pi$\\
SUCH THAT $\piszczalkaN{x}{\pi}{x}{E}$\\
HAVING $\lOne[\pi] - \mathrm{Nodes}[y] = 0$
\
}
\
Note the use of an existantial variable $y$ and how we count the number of all nodes in a graph with the $\mathrm{Nodes}$ labelling.

Now, we construct the query $Q_{\textrm{unique}}(\pi)$.
First, we express $Q_{\textrm{repeats}}(\pi)$ that holds for paths with some node occuring at least twice.

\qbox{LET $\labelling{loop}(x, y) := x=y$ IN\\
\match PATHS $\pi$\\
SUCH THAT $\piszczalkaN{z}{\pi'}{z}{\labelling{loop}}$\\
HAVING $\labelling{loop}[\pi, \pi'] \geq 2$
}
It states that $\pi'$ consists of the same node $u$ repeated multiple times and we require that there are at least two positions $i$ in $\pi$ such that
$u = \pi'[i] = \pi[i]$.

Finally, as we assume that all \opra{} queries are closed under the complement, there exists a query $Q_{\textrm{unique}}(\pi)$ in \opra{} that is the complement of $Q_{\textrm{repeats}}(\pi)$.
\end{proof}

Here are some examples of employing the closure properties. To check whether a given graph is a \emph{directed acyclic graph}, we have to check that the graph has no cycle.
Instead, we can check whether the graph has a cycle using the following query and then complement this query:

\qbox{\match ()
SUCH THAT $\piszczalkaN{x}{\pi}{x}{E}$
WHERE $\nc{\top}\nc{\top}\nc{\top}^*(\pi)$
}

The above query is Boolean, i.e., it has no free variables, so it is considered over an empty tuple, $()$.
This query checks whether there exists a path with the same initial and final nodes of length at least $2$, i.e., a cycle.

Finally, we can write a query that asks whether there is a unique path between $x$ and $y$.
First, the following query asks for nodes $x,y$ connected with at least two different paths.

\qbox{LET $\labelling{loop}(x, y) := \labelling{AND}(x=y, x \neq \padding)$ IN\\
\match NODES x, y\\
SUCH THAT $\piszczalkaN{x}{\pi}{y}{E} $ AND $ \piszczalkaN{x}{\pi'}{y}{E}$\\
HAVING $\labelling{loop}[\pi, \pi'] \geq  1$
}

Next, we take the complement of the above query and intersect that complement with the following query stating that there is at least one path from $x$ to $y$:

\qbox{\match NODES x, y\\
SUCH THAT $\piszczalkaN{x}{\pi}{y}{E}$
}

 \section{Expressive power}%
\label{sec:ep}
To understand the expressive power of \opra{}, we compare it with other languages. Let us first mention that \opra{} expresses all SQL queries over relational databases (subject to technical details arising from types and the fact that SQL can return an ordered list with repetitions). Most of the main ingredients of the proof of this claim are presented in Theorem~\ref{t:closure}, where we have shown the closure properties of \opra{}. We skip the proof because it provides little insight into what we are really interested in --- graph-oriented properties. Instead, we compare \opra{} with a well-known graph query language \ECRPQ{} and its extension with linear constraints  (\ECRPQLC)~\cite{Barcelo+12}. We prove the results depicted in Figure~\ref{fig:diagram}: that \pr{} subsumes \ECRPQ and \pra{} subsumes \ECRPQLC.
The strength of \ECRPQ{} comes from the possibility of comparing properties of paths that are expressible by synchronized regular automata. Nevertheless, \ECRPQ{} cannot deal with data values. Therefore, in the final part we show that \opra{} subsumes Regular Queries with Memory (RQM)\cite{Libkin+16} over graphs with integer data values.
We conclude with a short discussion on additional expressive power of \opra{} over \pra{}.

An \emph{\ECRPQ graph}~\cite{Barcelo+12} is a tuple $\tuple{V, E}$,
where $V$ is a finite set of nodes, and
$E \subseteq V \times \Sigma \times V$ is a set of edges labelled by a finite alphabet $\Sigma$.
A path in an \ECRPQ graph $G$ is a sequence of interleaved nodes and edge labels $v_0 e_0 v_1 \ldots v_k$ such that for every $i<k$ we have $E(v_{i}, e_{i}, v_{i+1})$.
The difference between \ECRPQ graphs and our graphs is mostly syntactical, yet obscures the close relationship between \ECRPQ and \pr.
To overcome this problem, we define the \emph{standard embedding}, which is a natural transformation of \ECRPQ graphs to graphs. The main idea is to replace paths of the form $v_0e_0v_1e_1 \dots v_n$ with paths of the form $(v_0, e_0)(v_1,e_1) \dots (v_{n-1}, e_{n-1}) (v_n, \padding)$.

The standard embedding of an \ECRPQ graph
$G = \tuple{V, E}$ over $\Sigma = \set{b_1, \dots, b_k}$ is a graph $\emb{G}$ whose set of nodes is $\emb{V} = V \times \Sigma_\padding$, where $\Sigma_\padding = \Sigma \cup \set{\padding}$. The graph is equipped with a binary Boolean labelling $\emb{E}$ encoding the edge relation: $\emb{E}((v, a), (v', a'))=1$ if and only if $E(v, a, v')$, and $|\Sigma_\padding|$ unary Boolean labellings $\lambda_b$ encoding the edge labels: $\lambda_b(v_1, a)=1$ if and only if $a = b$. To deal with variables that occur multiple times in path constraints (e.g.\ $x$ in $\piszczalka{x}{\pi}{x}$), we need an additional Boolean binary labelling $\sim$ that ties nodes representing the same node in $G$: $\sim((v,a),(v',a'))=1$ if and only if $v = v'$, for every $(v,a), (v', a') \in \emb{V}$. We say that a node $v$ \emph{corresponds} to the node $\emb{v}=(v,\padding)$, and that
a path $p=v_1 e_1 v_2 \dots v_n$ \emph{corresponds} to the path $\emb{p}=(v_1, e_1) \dots (v_{n-1}, e_{n-1}) (v_n, \padding)$.

\subsection{Extended conjunctive regular path queries (\ECRPQs)}
An \ECRPQ $Q(\vec{x},\vec{\pi})$ over $\Sigma$ is a conjunction of \emph{path constraints} of the form $\piszczalka{x_i}{\pi_k}{x_j}$ and
\emph{regular-relation constraints} of the form $\reg{R}(\pi_{i_1}, \ldots, \pi_{i_n})$, where $x_i,x_j$ are node variables, $\pi_k, \pi_{i_1}, \ldots, \pi_{i_n}$ are path variables,
and $\reg{R}$ is a regular expression defining an $n$-ary regular relation over $\Sigma$.
An \ECRPQ $Q(\vec{x},\vec{\pi})$ can contain other node and path variables beside those listed among $\vec{x}$ or $\vec{\pi}$; the remaining nodes and path variables are existentially quantified.

The language of \ECRPQs is based on the notion of \emph{regular relations}.
An $n$-ary relation $R$ on $\Sigma^*$ is regular if there is a regular expression $\reg{R}$ over the alphabet ${(\Sigma \cup \{\padding\})}^n$
such that for all words $w_1, \ldots, w_n \in \Sigma^*$ we have
$(w_1, \ldots, w_n) \in R$ if and only if $\comb{w_1, \ldots, w_n} \in L({\reg{R}})$ (the notion $\comb{p_1, \dots, p_k}$ has been introduced in Section~\ref{sec:opra} to define the semantics of regular constraints).
Note that we use the symbol $\padding$ to deal with the differences of paths' lengths, and hence, we need regular expressions over the alphabet ${(\Sigma \cup \padding)}^n$ to define $n$-ary relations over $\Sigma$.

The semantics of \ECRPQs is defined with respect to an \ECRPQ graph $G$ and an instantiation of all node and path variables $\nu$, i.e.,
for a nodes $\vec{v}$ of $G$ and paths $\vec{p}$ in $G$, we have that $Q(\vec{v},\vec{p})$ holds in $G$ if and only if there is an instantiation $\nu$ of nodes and path variables, which
is consistent with $\vec{v}$ and $\vec{p}$ on free nodes and respective path variables and such that all constraints of $Q(\vec{x},\vec{\pi})$ are satisfied.
A constraint $\piszczalka{x_i}{\pi_k}{x_j}$ is satisfied by $\nu$ if $\nu(\pi_k)$ is a path from $\nu(x_i)$ to $\nu(x_j)$; the semantics is the same as that of $\piszczalkaN{x_i}{\pi_k}{x_j}{E}$ in \pr{}.
The \ECRPQ graph $G$ and $\nu$ satisfy $\reg{R}(\pi_{i_1}, \ldots, \pi_{i_n})$
 if  and only if the sequences of labels of paths $\nu(\pi_{i_1}), \ldots, \nu(\pi_{i_n})$ belong to the relation defined by $\reg{R}$.

\ECRPQs are defined in a similar way to \pr{} queries.
However, regular-relation constraints in \ECRPQs and regular constraints are different.
In the case of a single path, regular-relation constraints specify regular languages of labels, while
regular constraints specify regular languages of node constraints, which are supposed to match the path.
Node constraints can express that a given node has a specific label and hence regular constraints (over a single path) can specify that a path has the sequence of labels
from a given regular language. The same reasoning works for multiple paths and it shows that regular constraints from \pr{} polynomially subsume regular-relation constraints from \ECRPQs.

A query $Q_1$ on \ECRPQ graphs is \emph{se-equivalent} to a query $Q_2$ on graphs if for all \ECRPQ graphs $G$, nodes $\vec{v}$ and paths $\vec{p}$, we have $Q_1(\vec{v}, \vec{p})$ holds in $G$ if and only if
$Q_2(\emb{\vec{v}}, \emb{\vec{p}})$ holds in $\emb{G}$.
A query language $\lang$ on graphs \emph{subsumes} a query language $\lang'$ on \ECRPQ graphs if for every query in $\lang'$ there exists a se-equivalent $\lang$ query.
Moreover, $\lang$ \emph{polynomially subsumes} $\lang'$ if every query in $\lang$ can be transformed to a query in $\lang'$ and the underlying transformation of queries is effective and takes polynomial time.

\begin{thm}%
\label{th:subsumesECRPQ}
(1) \pr{} polynomially subsumes \ECRPQ.
(2) There is a \pr{} query $Q$ with no \ECRPQ query $Q'$ se-equivalent to $Q$.
\end{thm}
\begin{proof}
\noindent \textbf{(1):} \ECRPQs consist of two types of constraints. Path constraints $\piszczalka{x_i}{\pi_k}{x_j}$ of \ECRPQ have similar semantics to
path constraints $\piszczalkaN{x_i}{\pi_k}{x_j}{E}$ in \pr{}, but there is a subtle difference arising from different path representation.
For example, if we take an \ECRPQ graph with an edge $(v, a, v)$, then $\piszczalka{x}{\pi}{x}$ should be satisfied by $\pi = v a v$. However, then $\emb{\pi}$ becomes $(v, a)(v, \padding)$
that has different endpoints. Therefore we do as follows.
For each $\piszczalka{x_i}{\pi_k}{x_j}$ we use a fresh variable $x_i'$.
The translation now consists of a path constraint $\piszczalkaN{x_i'}{\pi_k}{x_j}{\emb{E}}$
and three regular constraints: $\nc{\lambda_{\padding}(\z{x_i}{+0})=1}$, $\nc{\lambda_\padding(\z{x_j}{+0})=1}$ and $\nc{\sim(\z{x_i}{+0}, \z{x_i'}{+0})=1}$
Note that in the translation of $\piszczalka{x}{\pi}{x}$ the last of the regular constraints has the form $\nc{\sim(\z{x}{+0}, \z{x'}{+0})=1}$.

The regular-relation constraints of \ECRPQs are basically regular expressions over the alphabet ${(\Sigma \cup \set{\padding})}^n$.
In \pr{}, any letter $(a_1, \dots, a_n) \in {(\Sigma \cup \set{\padding})}^n$
can be expressed as the node constraint $\nc{\lambda_{a_1}(\z{\pi_1}{+0}) = 1 \wedge \ldots \wedge \lambda_{a_n}(\z{\pi_n}{+0}) = 1}$
referring to the current positions on the respective path variables of \ECRPQ.
 This can be extended to a translation of all regular-relation constraints in a straightforward way.
Hence \pr{} polynomially subsumes \ECRPQ.

\noindent \textbf{(2):}
Consider the following \pr{} query $Q_{b}$:
\begin{figure}
\begin{tikzpicture}

\begin{scope}
\node[circle,draw,minimum size=0.5cm] (V0) at (0,0) {$v_0$};
\node[circle,draw,minimum size=0.5cm] (V1) at (1.5,0) {$v_1$};
\node[] (ddd) at (3,0) {$\ldots$};
\node[circle,draw,minimum size=0.5cm] (VK2) at (4.5,0) {$v_{m}$};

\draw[->,bend left] (V0) to node[above] {a} (V1);
\draw[->,bend left] (V1) to node[above] {a} (ddd);
\draw[->,bend left] (ddd) to node[above] {a} (VK2);

\draw[->,bend left] (V1) to node[below] {a} (V0);
\draw[->,bend left] (ddd) to node[below] {a} (V1);
\draw[->,bend left] (VK2) to node[below] {a} (ddd);

\node[] at (1.5,-1.2) {$G$};

\end{scope}

\begin{scope}[xshift=7cm]
\node[circle,draw,minimum size=0.5cm] (U0) at (0,0) {$u_0$};
\node[circle,draw,minimum size=0.5cm] (U1) at (1.5,0.8) {$u_1$};
\node[circle,draw,minimum size=0.5cm] (U11) at (1.5,-0.8) {$u_1$};
\node[circle,draw,minimum size=0.5cm] (U2) at (3,0) {$u_2$};

\draw[->,loop above] (U0) to node[left] {a} (U0);
\draw[->,loop above] (U2) to node[right] {a} (U2);

\draw[->] (U0) to node[above] {a} (U1);
\draw[->] (U1) to node[above] {a} (U2);

\draw[->] (U2) to node[above] {a} (U11);
\draw[->] (U11) to node[above] {a} (U0);

\node[] at (0.7,-1.2) {$G'$};

\end{scope}
\end{tikzpicture}
\caption{The graphs $G$ and $G'$.}%
\label{fig:graphs}
\end{figure}
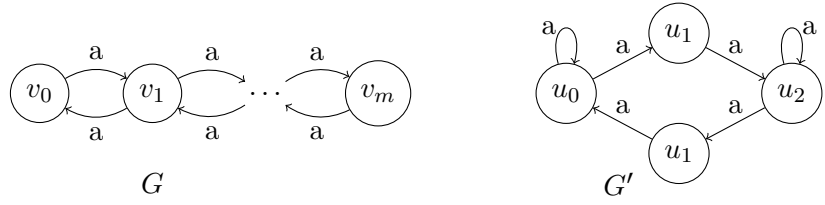

\qbox{\match NODES $x, y$
SUCH THAT $\piszczalkaN{x}{\pi}{y}{E}$
WHERE $\nc{E(\z{\pi}{+1}, \z{\pi}{+0})}^*\nc{\top}$}
which holds on nodes $x, y$ such that there is a bidirectional path between $x$ and $y$. We claim that this query is not expressible in \ECRPQ.

Suppose that there is an \ECRPQ $Q'(x,y)$ that holds if and only if there is a bidirectional path between $x$ and $y$.
Let $m$ be the number of all node variables in $Q'$.
Consider the graphs $G, G'$ depicted in Figure~\ref{fig:graphs}. We show that if $Q'(v_0, v_{m})$ holds in $G$, then $Q'(u_0, u_2)$ holds in $G'$.
Indeed, consider a node variable instantiation $\nu$ of $Q'(v_0, v_{m})$ in $G$. There is some node $v_j$ of $G$, which is not referred by any node variable.
We define a node variable instantiation $\nu'$ of $Q'(u_0, u_2)$ such that $\nu'(x) = u_0$ if $\nu(x) = v_i$ and $i <j$, and  $\nu'(x) = u_2$ otherwise.

Observe that if there is a path in $G$ of length $l$ between two nodes $\nu(x)$ and $\nu(y)$, then there is a path of the same length between $\nu'(x)$ and  $\nu'(y)$ in $G'$.
Similarly, if there is a path in $G$ of length $l$ from (resp., to) $\nu(x)$, then there is a path of the same length in $G'$ from (resp., to) $\nu'(x)$.
It follows that we can extend the instantiation $\nu'$ to path variables such that paths in $\nu$ and $\nu'$ have the same endpoints among instances of node variables and the same lengths.
Therefore, all constraints from $Q'(u_0, u_2)$ of the form $\piszczalka{x_i}{\pi_k}{x_j}$ are satisfied in $G'$ under $\nu'$.
Finally, since for every path variable $\nu(\pi)$ and $\nu'(\pi)$ have the same length and $a$ is the only label, all regular constraints of $Q'(u_0, u_2)$ are satisfied as well.
Hence, we have that $Q'(u_0, u_2)$ holds in $G'$, but there is no bidirectional path between $u_0$ and $u_2$.
\end{proof}

\subsection{\ECRPQs with linear constraints}
\ECRPQLC~\cite{Barcelo+12} is an extension of \ECRPQ with \emph{linear constrains}, expressing that a given vector of paths $\vec{\pi}$ satisfying a given \ECRPQ query satisfies linear inequalities,
which specify the multiplicity of edge labels in various components of $\vec{\pi}$.
Formally, a~\emph{linear constraint} is given by $h>0$, a $h \times (|\Sigma| \cdot n)$  matrix $A$ with integer coefficient and a vector $\vec{b} \in \Z^{h}$.
An instantiation $\nu$ of $\vec{\pi}$ (of length $n$) satisfies this constraint if $A \vec{l} \leq \vec{b}$ holds for the vector
$\vec{l} = (l_{1,1}, \ldots, l_{|\Sigma|,1}, l_{1,2}, \ldots, l_{|\Sigma|, n})$, where $l_{j,i}$ is the number of occurrences of the $j$-th edge label
in $\nu(\vec{p}[i])$.
In \ECRPQLC, we require a tuple of paths to satisfy both the ECRPQ part and the linear constraints.

Linear constraints can be expressed by arithmetical constraints of \pra{}.
This and Theorem~\ref{th:subsumesECRPQ} imply that \pra{} polynomially subsumes \ECRPQLC.
Still, linear constraints do not help with expressing structural graph properties.
In particular, \ECRPQLC{} does not express the query ``x and y are connected with a bidirectional path'', which is expressible in \pr{}.
Nevertheless, there are \ECRPQLC{} queries not expressible in \pr.
In consequence,  \ECRPQLC{} and \pr{} are incomparable and we have the following.

\begin{thm}\label{th:subsumesECRPQLC}%
\label{th:incomparable}
(1) \pra{} polynomially subsumes \ECRPQLC. (2) There is a \pr{} query $Q$ with no \ECRPQLC{} query $Q'$ se-equivalent to $Q$.
(3) There is an \ECRPQLC{} query $Q$ with no \pr{} query $Q'$ se-equivalent to $Q$.
\end{thm}

\begin{proof}
\noindent \textbf{(1):}
Language \pra{} can express all \ECRPQs with linear constraints.
Consider a query $Q$ of \ECRPQLC.
First, due to Theorem~\ref{th:subsumesECRPQ}, \pr{} polynomially subsumes ECRPQs, and hence there is a \pr{} query $\varphi_E$ corresponding to the \ECRPQ part of $Q$.
Second, we can express each $l_{j,i}$ by the arithmetical atom $\Lambda_{j,i} = \lambda_{b_j}[\pi_i]$.
Then, for $k = 1, \ldots, h$, the arithmetical constraint $\varphi_A^k$ corresponding to the $k-th$ row of $A$ can be constructed as the product of the row of $A$ and all the atoms $\Lambda_{j,i}$ compared to the $k$-th element of the vector $b$.
Thus, $\varphi_A$, defined as the conjunction of all $\varphi_A^k$, corresponds to the linear constraints $A \vec{l} \leq \vec{b}$ of $Q$. Finally, we define the \pra{} query se-equivalent to $Q$ as the conjunction of $\varphi_E$ and $\varphi_A$.

\noindent \textbf{(2):}
Consider the \pr{} query $Q_{b}$ presented in the proof of (2) in Theorem~\ref{th:subsumesECRPQ}.
The argument from (2) in Theorem~\ref{th:subsumesECRPQ} straightforwardly extends to \ECRPQLC.
We assume towards a contradiction that there is an \ECRPQLC $Q'(x,y)$ that holds if there is a bidirectional path between $x$ and $y$.
Let $k$ be the number of all node variables in $Q'$.
Then, we proceed as in the proof of (2) in Theorem~\ref{th:subsumesECRPQ} to show that if $Q'(v_0, v_{k})$ is satisfied under some instantiation $\nu$ in $G$ (depicted in Figure~\ref{fig:graphs}), then
we can define the corresponding instantiation $\nu'$ in $G'$ (depicted in Figure~\ref{fig:graphs}) such that
paths in $\nu$ and $\nu'$ have the same endpoints among instances of node variables and the same lengths.
Therefore, the \ECRPQ part of $Q'(u_0, u_2)$ holds in $G'$ under $\nu'$.
Observe that the value of the vector $\vec{l}$ in the linear constraints of $Q'(v_0, v_{k})$ under $\nu$ in $G$ is the same as the value of $\vec{l}$ under $\nu'$ in $G'$.
Thus, the linear constraints of $Q'$ are satisfied in $G'$ under $\nu'$.
Hence, we have that $Q'(u_0, u_2)$ holds in $G'$, but there is no bidirectional path between $u_0$ and $u_2$.

\noindent \textbf{(3):}
Consider the query $Q_{a=b}(\pi)$: ``a given path $\pi$ has the same number of edges labelled $a$ as edges labelled $b$''. We claim that $Q_{a=b}(\pi)$ is expressible in \ECRPQLC, whereas it is not expressible in \pr{}.

To see this, let us fix a graph $G$ consisting of a single state and two self-loops labelled by $a$ and $b$ respectively.
Consider the set of labellings of all paths satisfying $Q_{a=b}$. This set can be regarded as the language of
words over $\{a,b\}$ with the same number of $a$'s and $b$'s.
This language is not regular whereas for any \pr{} query $Q$, the language of labellings
of paths from $G$ satisfying $Q$ is regular.

Indeed, suppose that $Q$ has $k$ free path variables and no other path variables.
Observe that the language $L_Q$ of labellings of paths satisfying $Q$ is a regular language over ${\{a,b,\padding\}}^k$, i.e., $k$ element tuples over $\{a,b,\padding\}$.
Now, if $Q'$ is obtained from $Q$ by making some free path variables existentially quantified, then
$L_{Q'}$ is obtained from $L_{Q}$ by projecting out the components ${\{a,b,\padding\}}^k$ that correspond to existentially quantified variables.
Such an operation preserves regularity of the language; the language $L_{Q'}$ is still regular.
\end{proof}

Now we discuss why \opra{} is provably stronger than \pra{}.
\begin{rem}[\opra{} is stronger than \pra{}]%
\label{rem:OpraStronger}
The language \opra{} syntactically contains \pra{} and it strictly subsumes \pra{}.
Consider the property: ``an input graph is a directed acyclic graph (dag)''. \pra{} queries are monotonic and,
in consequence, no \pra{} query expresses the property.
On the other hand it is expressible in \opra{}. We can write a boolean \pr{} query $Q()$ that holds on input graphs with a cycle. From Theorem~\ref{t:closure} the complement of $Q()$ can be expressed in \opra{}.
\end{rem}

\subsection{Regular queries with memory (RQMs)}

Regular Query with Memory (RQM)~\cite{Libkin+16,LibkinTV15} is of the form $\piszczalka{x}{\pi}{y} \wedge \pi \in L(e)$, where $e$ is a regular expression with memory (REM)~\cite{Libkin+16,LibkinTV15}.
We refrain from presenting a formal definition of REMs as we do not use it below. Intuitively, REMs can store in a variable the data value at the current position and test its (dis)equality with other values already stored. RQMs are evaluated over \emph{data graphs}~\cite{Libkin+16}.
A data graph $G$ over a finite alphabet $\Sigma$ and countable infinite set $\mathcal{D}$ is a triple $(V, E, \rho)$, where $V$ is a finite set of nodes,
$E\subseteq V \times \Sigma \times V$ is a set of labelled edges; and $\rho: V \to \mathcal{D}$ is a function that assigns a data value to each node in $V$.
In this paper we assume that $\mathcal{D}$ is a set of integer numbers.
A path in a data graph $G$ is a sequence of interleaved nodes and edge labels $v_0 e_0 v_1 \ldots v_k$ such that for every $i<k$ we have $E(v_{i}, e_{i}, v_{i+1})$. REMs are evaluated over \emph{data paths}. Given a path $p = v_0 e_0 v_1 \ldots v_k$, a data path corresponding to $p$ is $\rho(v_0) e_0 \rho(v_1) \ldots \rho(v_k)$, i.e., a sequence of alternating data values and labels that starts and ends with data values.

Given a data graph $G$, the result of the RQM $\piszczalka{x}{\pi}{y} \wedge \pi \in L(e)$ on $G$ consists of pairs of nodes $(v, v')$ such that there is a data path $w$ from $v$ to $v'$ that belongs to $L(e)$.

In order to relate \pra{} to RQM we apply a natural transformation of data graphs to graphs.
We discussed it in Section~\ref{sec:prelim}, see the examples~\ref{fig:propertygraph} and~\ref{fig:map}.
The standard embedding of a data graph $G_{\mathrm{data}} = (V, E, \rho)$ is a graph $\demb{G}$
whose set of nodes is $\demb{V} = V \cup E$. $\demb{G}$ is equipped with the following labellings:
\begin{itemize}
\item a binary Boolean labelling $\demb{E}$ encoding the edge relation:
for each $v\in V$ and $e\in E$ we set $\demb{E}(v, e)=1$ if and only if there is $v'$ such that $e = (v, a, v') \in E$ and we set $\demb{E}(e, v)=1$ if and only if there is $v'$ such that $e = (v', e, v) \in E$;
\item a unary labelling $\lambda$ encoding data labelling of $G_{\mathrm{data}}$ defined as  $\lambda(v) = \rho(v)$ if $v\in V$ and $\lambda(v) = 0$ otherwise.
\item for each $b \in \Sigma$, a unary Boolean labelling $\lambda_b$ encoding the edge labels: for each $e = (v, a, v')\in E$ we set $\lambda_b(e)=1$ if $a = b$ and
$\lambda_b(e)=0$ otherwise.
\end{itemize}

\noindent
We say that a query $Q_1$ on data graphs is \emph{se-equivalent} to a query $Q_2$ on graphs if for all data graphs $G$ and nodes $v, v'$ the
query $Q_1(v, v')$ holds in $G$ if and only if $Q_2(v, v')$ holds in $\demb{G}$.

Similarly as before, a query language $\lang$ on graphs \emph{subsumes} a query language $\lang'$ on data graphs if for every query in $\lang'$ there exists a se-equivalent $\lang$ query.
Also, $\lang$ \emph{polynomially subsumes} $\lang'$ if every query in $\lang'$ can be transformed to a query in $\lang$ and the transformation is effective and can be computed in polynomial time.

Now, we would like to prove that \pra{} subsumes RQM. To make the proof easier we introduce an intermediate step through \emph{regular data path queries (RDPQ)}~\cite{Libkin+16}.
RDPQs are automata-based formalism and, like RQMs, define pairs of nodes of data graphs.
In the proof, given a RQM $Q$ we express it by RDPQ $Q_1$ and then we construct a se-equivalent \pra{} query.

Now we formally define RDPQs~\cite{Libkin+16}.
An RDPQ is of the form $\piszczalka{x}{\pi}{y} \wedge \pi \in L(\mathcal{A})$, where $\mathcal{A}$ is a \emph{Register Data Path Automaton (RDPA)}.
RDPAs, similarly as REMs, are evaluated over data paths.
In order to compare the data values, RDPA use Boolean combinations of the conditions of the form \(x_i^=\), \( x_i^\neq \), \(z^=\) and \(z^\neq \),
where each variable $x_i$ refers to the $i$-th register and $z$ is a data value from $\mathcal{D}$ (a constant).
Let $\mathcal{C}_k$ be the set of all such conditions over $k$ registers and their Boolean combinations. Each of the registers store either a data value or a special value $\bot$ that means that the register has not been assigned yet.
Semantics of the conditions is defined with respect to a (current) data value $d$ and a valuation of registers $\tau = (d_1, \ldots, d_k) \in {(\mathcal{D} \cup {\bot})}^k$ in a natural way:
for each $\otimes \in \set{=, \neq}$ and each $i \in \{1, \ldots, k\}$ we define
$d, \tau \models x_i^\otimes$ iff $d \otimes d_i$, and $d, \tau \models z_i^\otimes$ iff $d \otimes z$. In th sequel, given a register valuation $\tau = (d_1, \ldots, d_k)$ we will write $\tau(i)$ for $d_i$, the $i$th element of the tuple $\tau$.

A $k$-register RDPA~\cite{Libkin+16} consists of a finite set of \emph{word states} $Q_w$, a finite set of \emph{data states} $Q_d$, an initial state $q_0\in Q_d$, a set of final states $F \subseteq Q_w$ and two transition relations $\delta_w$, $\delta_d$ such that:
\begin{itemize}
\item $\delta_w \subseteq Q_w \times \Sigma \times Q_d$ is the word transition relation;
\item $\delta_d \subseteq Q_d \times \mathcal{C}_k \times 2^{\set{1, \ldots, k}} \times Q_w$ is the data transition relation.
\end{itemize}
Given a data path $w = d_0 e_1 d_2 \ldots d_l$,  where each $d_i \in \mathcal{D}$ and each $e_i \in \Sigma$, a \emph{computation} of $\mathcal{A}$ on $w$ is a sequence of tuples $(0,q^0, \tau^0), \dots, (l+1, q^{l+1}, \tau^{l+1})$, where $q^0=q_0$, $\tau^0= \bot^k$ and:
\begin{itemize}
\item for each even $j$ there is a transition $(q^j, c, I, q^{j+1}) \in \delta_d$ such that
$d_j, \tau^j \models c$ and for each $i \in \{1, \ldots, k\}$ the value $\tau^{j+1}(i)$ is equal to $d_j$ if $i \in I$ and to $\tau^j(i)$ otherwise.
\item for each odd $j$, there is a transition $(q^j, e_j, q^{j+1})\in\delta_w$ and $\tau^{j+1}=\tau^j$.
\end{itemize}

\noindent
A data path $w$ is \emph{accepted} by $\mathcal{A}$ if $\mathcal{A}$ has a computation on $w$ that ends in a configuration containing a final state.
Given a data graph $G$, the result of the RDPQ $\piszczalka{x}{\pi}{y} \wedge \pi \in L(\mathcal{A})$ on $G$ consists of pairs of nodes $(v, v')$ such that there is a data path $w$ from $v$ to $v'$ that is accepted by $\mathcal{A}$.

Each RQM can be expressed with a RDPQ. This is because for each REM with $k$ variables one can construct in PTIME a RDPA with $k$ registers that accepts the same language of data paths~\cite[Proposition 3.13]{Libkin+16} and~\cite[Theorem 4.4 ]{LibkinTV15}.

We show that \pra{} subsume RQM. The transformation from an RQM to a \pra{} involves a single exponential blow-up.

\begin{thm}\label{th:subsumeRQM}
(1) \pra{} subsumes RQM. (2) There is an \opra{} query $Q$ with no RQM query $Q'$ equivalent to $Q$.
\end{thm}
\begin{proof}
Let $Q$ be an RDPQ of the form $\piszczalka{x}{\pi}{y} \wedge \pi \in L(\mathcal{A})$, where $\mathcal{A}$ is a RDPA.
We will construct an se-equivalent \pra{} query as follows.
First, assume that $\mathcal{A}$ contains exactly one final state. We explain later what to do when this is not true.
We construct an \emph{intermediate Path Automaton (iPA)} $\mathcal{A'}$ that is equivalent to $\mathcal{A}$ in the following sense:
given a data graph $G$, nodes $v, v'$ of $G$ the RDPA $\A$ accepts a data path from $v$ to $v'$ if and only if $A'$ accepts a path from $v$ to $v'$ in $\demb{G}$.
 The iPA is an automaton where transitions are labelled by regular constraints from \pra.

iPA process tuples of paths of a graph (like \pra) rather than data paths of a data graph (as in the case of RDPA).
A tuple of paths includes additional paths that store values during computation just like registers do in the case of RDPA.
Therefore, iPA do not have registers explicitly and whenever we mention \emph{registers} in the context of iPA we mean the mechanism to store values in additional paths.
Moreover, the stored values are nodes of a graph as in \pra{} queries and not data values as in RDPQ.

Formally, a $k$-register iPA consists of a set of states $Q$, a single initial state $q_I$, a single final state $q_F$, and a transition relation $\delta \subseteq Q \times \verb+Regular_constraints+ \times Q$.
The regular constraints in a $k$-register iPA use a path variable $\pi_0$ corresponding to an input path and $k$ existentially quantified, additional path variables $\pi_1, \ldots, \pi_k$ to store the values of the $k$ registers.

In what follows, for a path $p$, by $p[i,j]$ we denote the fragment of $p$ that starts at position $i$ and ends at the position $j$.
e.g.,
for $p = v_1 e_2 v_3 e_4 v_5$, the expression $p[3, 5]$ denotes $v_3 e_4 v_5$.

We say that for a given graph $G$, an iPA $\A'$ accepts a path $p_0$ if there are paths $p_1, \ldots, p_k$, each of the same length as $p_0$,
there are: a number $n \in \N$, positions $0 = i_0 < i_1 < i_2 < \cdots < i_{n-1} \leq i_{n} = |p_0|$ and a sequence of states $q_0, \ldots, q_n$
such that
\begin{itemize}
\item $q_0$ is the initial state, $q_n$ is the final state, and
\item for each $j\in\{ 0, \ldots, n-1\}$
there is a transition $(q_j, R(\pi_0, \ldots, \pi_k), q_{j+1}) \in \delta$ such that
\begin{itemize}
\item
 for $j<n-1$,  the paths $p_0[i_j,i_{j+1}], \ldots, p_k[i_j,i_{j+1}]$ satisfy the regular constraint $R \cdot \nc{\top}$, i.e.,~$R\cdot \nc{\top}(p_0[i_j,i_{j+1}], \ldots, p_k[i_j,i_{j+1}])$ holds.
\item for $j = n-1$,  the paths $p_0[i_j,i_{j+1}], \ldots, p_k[i_j,i_{j+1}]$ satisfy the regular constraint $R$
\end{itemize}
\end{itemize}
Note that for $j\in\{ 1, \ldots, n-1\}$ the positions $i_j$ on the paths $p_0, \ldots, p_k$ are processed twice, first by the always satisfied node constraint $\nc{\top}$ and only then by the actual transition. We apply this trick to be able to refer to the next position on the paths even if the current position is the last one in the current section. In other words,
for $0\leq j \leq n-2$ we include the position $i_{j+1}$ in the fragments $p_i[i_j,i_{j+1}]$ only to be able to refer to the next position while being at the position $i_{j+1}-1$.

Now, given a $k$-register RDPA $\A = (Q_w, Q_d, q_0, F, \delta_w, \delta_d)$ we define an equivalent $k$-register iPA $\A' = (Q, q_I, q_F, \delta)$.
Let $Q = Q_w \cup Q_d$. We define $q_I$ to be $q_0$ and $q_F$ to be the only state in $F$.

We now define a transition relation $\delta$.

For each transition $(q, a, q')\in \delta_w$ we define a new transition $(q, R, q')\in \delta$, where the regular constraint $R=\nc{\lambda_a(\z{\pi_0}{+0})=1 \wedge \demb{E}(\z{\pi_0}{+0}, \z{\pi_0}{+1})=1 \wedge \bigwedge_{i\in \{1, \ldots, k\}} \lambda(\z{\pi_i}{+1})=\lambda(\z{\pi_i}{+0})}$ ensures that
\begin{itemize}
\item the current node on $\pi_0$ corresponds to an edge in the graph that is labelled with $a$;
\item the edge goes to the next node on $\pi_0$;
\item for each path $\pi_1, \ldots, \pi_k$ the next value is the same as the current value (i.e., the registers values do not change).
\end{itemize}
Note that we can safely refer to the next position on paths $\pi_i$ because $q'$ is not the final state as $q_F \in Q_w$ and $\delta_w \subseteq Q_w \times \Sigma \times Q_d$.

For each transition $(q, c, I, q') \in \delta_d$, with a condition $c$ and an update set $I$, we put $(q, R, q')$ in $\delta$, where the regular constraint $R$ ensures that
\begin{enumerate}
\item there is an edge between the current and the next node on $\pi_0$ in $G$;
\item
 the condition $c$ is satisfied assuming for each $i \in \set{1, \ldots, k}$ the value of the register $i$ is the current value of the path $\pi_i$;
\item
 unless $q'$ is the final state, for each $i\in I$ the next value of the path $\pi_i$ is set to the current value of the input path $\pi_0$;
\item
 unless $q'$ is the final state, for each $i\in \{1,\ldots,k\}\setminus I$ the next value of the path $\pi_i$ is the same as the current value of $\pi_i$.
\end{enumerate}

\noindent
We express the condition $c \in \mathcal{C}_k$ as the Boolean combination of node constraints, denoted by $N(c)$, by replacing each $x_i^\otimes$ by $\nc{\lambda(\z{\pi_i}{+0}) \otimes \lambda(\z{\pi_0}{+0})}$ and $z^\otimes$ by $\nc{z\otimes \lambda(\z{\pi_0}{+0})}$, for $\otimes \in \set{=, \neq}$. Recall that $\lambda$ is the labelling encoding the data values of the nodes.

We cannot include $N(c)$ directly in a regular constraint because regular constraints are conjunctions of node constraints.
Hence, we transform $N(c)$ to DNF. Then we remove all negations swapping $=$ and $\neq$ in $x_i^\otimes$ and $z^\otimes$ as required.
 Denote the resulting expression by $N = \bigvee_{l} N_l$.

If $q'$ is not the final state then for each of the conjunctions $N_l$ we define a regular constraint $R_l$ as
\[\nc{\demb{E}(\z{\pi_0}{+0}, \z{\pi_0}{+1})=1 \wedge N_l \wedge \bigwedge_{i\in I} \lambda(\z{\pi_i}{+1})=\lambda(\z{\pi_0}{+0}) \wedge \bigwedge_{i\in \{1, \ldots, k\}\setminus I} \lambda(\z{\pi_i}{+1})=\lambda(\z{\pi_i}{+0})}\]

Otherwise $R_l$ is defined as $\nc{N_l}$.
Finally, we define $R$ as $R_1 {+} \cdots {+} R_s$, where $N =  N_1 \vee \ldots N_s$.
This finishes the construction of the automaton $\A'$.
The automaton $\A'$ has the same number of states as $\A$, but its size may be exponential in $|\A|$ due to transformation from an arbitrary Boolean combination  of node constraints to a DNF. This blow-up can be avoided if we allow \opra{} queries, in which we can define a new labelling corresponding to any Boolean function, and then use this labelling to express a Boolean combination of node constraints.

To end the proof of (1) we use the standard state removal method of converting an NFA to a regular expression. This removes all the states of $\A'$ except $q_I$ and $q_F$.
 Then we use the same techniques as when removing states to define from the remaining transitions (i.e., the loops on $q_I$, on $q_F$, and the transition from $q_I$ to $q_F$ and back) the single transition $t = (q_I, R, q_F)$ with the regular constraint $R$ such that $R$ describes equivalently all possible paths from $q_I$ and $q_F$.
We use $R$ to define \pra{} query  \q{SELECT  NODES x, y WHERE $R$} that is se-equivalent to the original RDPQ query $Q$.
The transformation from a $\A'$ to $R$ is exponential in the number of states of $\A'$, which is the same as for $\A$, and hence
the whole construction results in a single exponential blow-up.

If the RDPA $\mathcal{A}$ has more than one final state in $F$ we repeat the above construction $|F|$ times.
Namely, for each $q \in F$ we construct $\mathcal{A}_q$ which is identical as $\mathcal{A}$ but with $q$ as its only final state.
This way for each $q \in F$ we obtain a regular constraint $R_q$ and then we set $R$ as the alternation (+) of the regular constraints $R_q$ for all  $q \in F$.

In order to prove (2) it is enough to note that \opra{} can express the query ``an input graph is a dag'' as we discuss in Remark~\ref{rem:OpraStronger}. Clearly, RQMs are monotonic (i.e., if an RQM query holds in a graph $G$ then it holds in any $G'$ containing $G$) and cannot express this query.
\end{proof}

 \section{The query evaluation problem}\label{sec:summing}

\newcommand{\QTypeNode}{\underline{V}}
\newcommand{\QTypeLabelling}{\underline{\lambda}}
\newcommand{\QTypeFirst}{\underline{S}}
\newcommand{\QTypeLast}{\underline{T}}

 The query evaluation    problem asks, given an \opra{} query $Q(\vec{x},\vec{\pi})$, a graph $G$, nodes $\vec{v}$ and paths $\vec{p}$ of $G$, whether $Q(\vec{v},\vec{p})$ holds in $G$.
  We are interested in the \emph{combined complexity} of the query evaluation problem, where the size of the input is the size of $G$, paths $\vec{p}$ and an input \opra{} query, and
  in the \emph{data complexity}, where a query is fixed and only $G$ and $\vec{p}$ are the input.

\Paragraph{Unary encoding of graph labels} To obtain the desired complexity results, we assume that the absolute values of the graph labels are polynomially bounded in the size of a graph, or equivalently that graph labels are encoded in unary.
This allows us to compute arithmetical relations on these labels in logarithmic space.
Without such a restriction, the data complexity of the query evaluation problem we study is \NP{}-hard by a straightforward reduction from the knapsack problem.

For the combined complexity, we assume that auxiliary labellings have a \emph{bounded depth} defined as follows.
For auxiliary labellings $O :=\; \definedBy{\lambda_1}{t_1}, \ldots, \definedBy{\lambda_n}{t_n}$, we say that $\lambda_i$ depends on $\lambda_j$ if $t_i$ refers to $\lambda_j$.
The relation \emph{depends on} defines a directed acyclic graph on $\lambda_1, \ldots, \lambda_n$ and we define the \emph{depth} of $O$ as the maximal length of a path in this acyclic graph.
An \opra{} query $Q$  := \q{LET $O$ IN $Q'$} is an \emph{\opra{} query of depth (at most) $s$}, denoted by \opra{}[s], if $s = 0$, and $Q$ is a \pra{} query, i.e., $O$ is empty, or $s > 0$,
$O$ has depth at most $s$ and all subqueries of $Q'$ are \opra{}[s-1] queries.

\begin{thm}%
\label{th:query-in-NL}
The following holds:
\begin{enumerate}
\item The data complexity of the query evaluation problem for \opra{} queries is \NL{}-complete.
\item The combined complexity of the query evaluation problem for \opra{} queries with a bounded depth of auxiliary labellings is \PSpace{}-complete.
\end{enumerate}
\end{thm}

\noindent
The lower bounds in Theorem~\ref{th:query-in-NL} holds even for \pr{}. Indeed, the \NL{}-hardness result in (1) of Theorem~\ref{th:query-in-NL} follows from \NL{}-hardness of the reachability problem, which can be expressed even in \pr{}.
The \PSpace{}-hardness result in (2) Theorem~\ref{th:query-in-NL} follows from \PSpace{}-hardness of the query evaluation problem for \ECRPQs~\cite{Barcelo+12} and polynomial-time translation from
\ECRPQs to \opra{}~(Theorem~\ref{th:subsumesECRPQ}).
The upper bound in Theorem~\ref{th:query-in-NL} follows directly from the following Lemma~\ref{l:induction}.

\begin{lem}%
\label{l:induction}
For every fixed $s \geq 0$, we have:
\begin{enumerate}[(1)]
\item Given a graph $G$, nodes $\vec{v}$ and paths $\vec{p}$ of $G$, and an \opra{}[s] query $Q(\vec{x},\vec{\pi})$,
we can decide whether $Q(\vec{v},\vec{p})$ holds in $G$ in non-deterministic polynomial space in $|Q|$ and non-deterministic logarithmic space in $|G|$.
\item Given a graph $G$, nodes $\vec{v}$ of $G$, and an \opra{}[s] query $Q(\vec{x},{\pi})$,
we can compute $\min_{\lambda, \pi} Q(\vec{v}, \pi)$ (resp., $\max_{\lambda, \pi} Q(\vec{v}, \pi)$) in  non-deterministic polynomial space in $|Q|$ and non-deterministic logarithmic space in $|G|$. The computed value is either polynomial in $|G|$ and exponential in $|Q|$, or $-\infty$ (resp., $\infty$).
\end{enumerate}
\end{lem}

\begin{rem}[Functions computed non-deterministically]\label{nondeterministicfuns}
In Lemma~\ref{l:induction} values $\min_{\lambda, \pi} Q(\vec{v}, \pi)$, $\max_{\lambda, \pi} Q(\vec{v}, \pi)$ are computed on a non-deterministic machine, which is a non-standard notion.
However, some subcomputations of of our non-deterministic decision procedure return values. Therefore, we adopt non-deterministic framework to functional problems as follows.
We say that a non-deterministic machine $M$ computes a function $f$, if
for every input $w$, (1)~$M$ can have multiple computations on $w$ each \emph{accepting} or \emph{rejection},
(2)~for all accepting computations of $M$ on $w$, it returns the correct value $f(w)$,
(3)~$M$ has at least one accepting computation on $w$.
\end{rem}

Note that when considering data complexity the query is fixed and hence its depth is bounded.

We first prove the upper bounds for \pra{} (i.e., for $s=0$), and then extend the results to \opra{}. The general idea of the proof is in a similar vein as the proof of the upper bound of ECRPQ. However, since our language is much more complex, we use more sophisticated, well-tailored tools.

\subsection{Language \pra{}}
Assume a \pra{} query

\qbox{$Q=$ \match NODES $\vec{x}$, PATHS $\vec{\pi}$ SUCH THAT $P$ WHERE $R$ HAVING $A$}
We prove the results in two steps.  First, we construct a Turing machine of a special kind (later on called \qam{}) that represents graphs, called \emph{answer graphs}.
Given a query $Q$ and a graph $G$, the answer graph is a graph with distinguished initial and final nodes such that every path from an initial node to a final node in this graph is an encoding of a vector of paths that satisfy constraints $P$ and $R$ of $Q$ in graph $G$
(for some instantiation of variables $\vec{x}$). The answer graph is augmented with the computed values of the expressions that appear in the arithmetical constraints $A$.
Thus, the query evaluation problem reduces to the existence of a path in the answer graph satisfying $A$. The instantiation of $\vec{x}$ can be inferred from the path.

Second, we prove that checking whether there is a path from an initial node to a final node in the answer graph that encodes a path in  $G$ satisfying $A$ can be done within desired complexity bounds.
However, the answer graph for $Q$ and $G$ has a polynomial size in $G$ and hence it cannot be explicitly constructed in logarithmic space.
We represent these graphs \emph{on-the-fly} using Query Applying Machines (\qam{}).
Such a representation allows us to construct the answer graph and check the existence of a path satisfying $A$ in non-deterministic logarithmic space in $G$.
\smallskip

The first step is an adaptation of the technique commonly used in the field, e.g., in~\cite{Barcelo+12,Grabon+16}. We encode vectors $(p_1,\dots, p_n)$ of paths of nodes from some $V$ as a single path $p_1 \otimes \ldots \otimes p_n$
over the product alphabet $V^n$ (shorter paths are padded with $\padding$).

\Paragraph{Answer graphs}
Consider a graph $G$ with nodes $V$, its paths $\vec{p}$ and an \opra{} query

\newcommand{\cqp}{C_Q^{\vec{p}}}
\noindent \qbox{$Q=$ \match NODES $\vec{x}$, PATHS $\vec{\pi}$ SUCH THAT $P$ WHERE $R_1, \ldots, R_p$ HAVING $\bigwedge_{i=1}^m A_i \leq c_i$}
with $|\vec{p}| = |\vec{\pi}|$ and existentially quantified path variables $\vec{\pi}'$.
Let $k={|\vec{\pi}|+|\vec{\pi}'|}$ be the number of path variables.
For every regular constraint $R_i$ in $Q$, we build an NFA $\aut_i$ with the set of states $C_i$ recognizing the language of $R_i$. We extend each $\aut_i$ with self-loops on all final states labelled by $\bot$ --- an auxiliary node constraint satisfied if all the input nodes are $\padding$.
We define $\cqp$ as $C_1 \times \cdots \times C_{p} \times \{1, \ldots, N, \infty\}$, where $N$ is the maximal length of paths in $\vec{p}$.

The \emph{answer graph} for $Q$ on $G$, $\vec{p}$ is a triple $(G', S, T)$, where $S, T\subseteq \cqp{} \times V^k$;
\begin{itemize}
\item $G'$ is a graph with nodes $\cqp{} \times V^k$, labelled by $\lambda_1, \ldots, \lambda_m, \lambda_E$;
\item for each $i\leq m$ and a node $(s,v_1, \dots, v_k)\in \cqp \times V^k$, the labelling $\lambda_i(s,v_1, \dots, v_k)$ is defined as the value of the arithmetical constraint $A_i$ over a vector of single-node paths $v_1$, \dots, $v_{k}$;
and
\item{}
paths $q_1, \dots, q_k$ s.t. $(q_1, \dots, q_{|\pi|}) = \vec{p}$ satisfy $P \land R$ if and only if there is $c$ such that the path $q=c \otimes q_1 \otimes \dots \otimes q_k$ from (a node in) $S$ to (a node in) $T$ is
such that $\lambda_E(v, v')=1$ for all consecutive $v$, $v'$ of $q$.
\end{itemize}
Intuitively, the labelling $\lambda_E$ can be defined in such a way that along paths of $G'$
the $V^k$-components of the nodes correctly encode paths of $Q$ satisfying the path constraints $P$ and
the $\cqp$ components store valid runs of automata corresponding to the regular constraints $R$.

\Paragraph{Query Applying Machine (\qam{}s)} Answer graphs are typically to big to be stored in an explicit way, but can be represented (on-the-fly) in logarithmic space.
To do so, we introduce a Query Applying Machine (\qam{}). \qam{} is a non-deterministic Turing Machine, which works in logarithmic space and only accepts inputs encoding tuples of the form $(G, t, w)$, where
$G$ is a graph, $t$ is a \emph{query type} consisting of symbols $\QTypeNode, \QTypeLabelling, \QTypeFirst, \QTypeLast$, and $w$ is the object being queried.

For a graph $G$ and $k\geq 0$, a \qam{} $\M$ \emph{gives} a graph $G^\M_k = (V, \lambda_1, \dots, \lambda_k, \lambda_E)$ and sets of nodes $S^\M_G$, $T^\M_G$ such that:
\begin{itemize}
\item ${V}$ consists of all the \emph{nodes} $v$ such that $\M$ accepts on $(G, \QTypeNode,  v)$,
\item ${\lambda_i}$ is such that $\lambda_i(\vec{v}) = n$ if and only if $\M$ accepts on $(G, \QTypeLabelling, (i, \vec{v}, n))$
\item ${\lambda_E}$ is such that $\lambda_E(\vec{v}) = n$ if and only if $\M$ accepts on $(G, \QTypeLabelling, (E, \vec{v}, n))$
\item $S_G^\M$ (resp., $T_G^\M$) consists of $v\in V$ such that $\M$ accepts on $(G,  \QTypeFirst, v)$ (resp., $(G, \QTypeLast, v)$).
\end{itemize}
For soundness, we require that for each $G$, $i$ and $\vec{v}$ there is exactly one $n$ such that $\M$ accepts on $(G, \QTypeLabelling, (i, \vec{v}, n))$ and, analogously, for each $G$, $\vec{v}$ there is exactly one $n$ such that $\M$ accepts on $(G, \QTypeLabelling, (E, \vec{v}, n))$.

Given a \pra{} query and its parameters, we can construct a QAM that gives on-the-fly answer graphs for this query:

\begin{lem}\label{l:pratoqam}
For a given query $Q$ and paths $\vec{p}$, we can construct in
polynomial time a \qam{} $\M^Q$ such that for every graph $G$, the machine $\M^Q$ gives an answer graph for $Q$ on $G, \vec{p}$.
\end{lem}

\begin{proof}
\newcommand{\idxInp}{k_1}
\newcommand{\idxEx}{k_2}

Consider a vector $\vec{p}$ of paths of $G$ and an \opra{} query

\noindent \qbox{$Q=$ \match NODES $\vec{x}$, PATHS $\vec{\pi}$ SUCH THAT $P_1, \ldots, P_l$ WHERE $R_1, \ldots, R_p$ HAVING $\bigwedge_{i=1}^m A_i \leq c_i$}
 with $|\vec{p}| = |\vec{\pi}|$ and existentially quantified path variables $\vec{\pi}'$.
Let $\idxInp = |\vec{\pi}|$, $\idxEx = |\vec{\pi}'|$ and $k=k_1+k_2$.
We discuss how each of the components of an answer graph for $Q$ on an input graph $G$ with nodes $V$ and $\vec{p}$ can be constructed.

\noindent \emph{The nodes.} The set of nodes is $\cqp{} \times V^k$, which can be recognized in logarithmic space (note that $\cqp$ and $k$ do not depend on the input of the \qam{}).

\noindent \emph{The labelling $\lambda_E$.}
Let $\vec{u}_1, \vec{u}_2$ be nodes of $G'$.
We put $\lambda_E(\vec{u}_1, \vec{u}_2) = 1$ if nodes $\vec{u}_1, \vec{u}_2$ are \emph{path consistent} and \emph{state consistent}, defined as follows, $\lambda_E(\vec{u}_1, \vec{u}_2) = 0$ otherwise.

\emph{Path consistency.}
Let $\vec{u}_1 = (c, v_1, \ldots, v_k)$ with $c = (s_1, \ldots, s_p, j)$, and let
 $\vec{u}_2 = (c', v_1', \ldots, v_k')$ with $c' = (s_1', \ldots, s_p', j')$.
 Path consistency  ensures that the first $\idxInp$ components of $V^k$ encode the input paths $\vec{p}$, paths that end satisfy path constraints, and paths that have terminated do not restart, i.e.,
\begin{itemize}
\item $j' = j+1$ if $j <N$,
\item $j' = \infty$ if $j \geq N$,
\item for each $i \in \{1, \dots, \idxInp\}$ we have $v_i = p_i[j]$ or $v_i = \padding$ and $j > |p_i|$ (in particular if $j = \infty$),
\item for each $i \in \{\idxInp+1,\dots,  k\}$, if $v_i \neq \padding$ and $v_i' = \padding$ ($v_i$ is the last node of $\pi_i$), then for every
path constraint $\piszczalka{x_s}{\pi_i}{x_t}$, we require $v_i = x_t$,
\item for each $i \in \{\idxInp+1, \dots,  k\}$
~if $v_i = \padding$, then $v_{i}' = \padding$.
\end{itemize}

\noindent \emph{State consistency.}
The state consistency ensures that the component $\cqp$ stores valid runs of automata for the regular constraints, i.e., that for each $i \in \{1, \ldots, p\}$, the automaton $\aut_i$ has a transition $(s_i, a_i, s_i')$, where $a_i$ is a node constraint, and $a_i$ is satisfied over the nodes of $v_1, v_1', \ldots, v_k, v_k'$ (we assume that $a_i$ selects from the list of all paths only the relevant paths listed in the regular constraint).

It is easy to check that given a graph $G$ and its two nodes $\vec{v}_1, \vec{v_2}$, we can decide in logarithmic space in $G$ whether
$\lambda_E(\vec{v}_1, \vec{v}_2)$ is $0$ or $1$.

\noindent \emph{The labellings $\lambda_1, \ldots, \lambda_m$.}  For $i \in \{1, \ldots, m\}$ we define $\lambda_i(c, v_1, \ldots, v_k)$ as the value of $A_i$ computed on the subset of $v_1, \ldots, v_k$ selected by $A_i$.
Since $A_i$ is a linear combination and each labelling of $G$ is given in unary, all labellings of $G'$ can be computed in logarithmic space in $G$.

\noindent \emph{The initial and final sets $S$ and $T$.}  The set $S$ consists of the nodes $(c, v_1, \ldots, v_k,1)$ such that
(1)~$c = (s_1, \ldots, s_p)$ and $s_i$ is an initial state of $\aut_i$ for $i \in \{1, \ldots, p\}$,
(2)~for every path constraint $\piszczalka{x_s}{\pi}{x_t}$, we require $v_i = x_s$, and
(3)~for every $i \in \{1, \ldots, \idxInp\}$ we have $v_i = p_i[1]$.
 The set $T$ consists of the nodes $(c, v_1, \ldots, v_k)$ such that
(1)~$c = (s_1, \ldots, s_p,\infty)$ and $s_i$ is a final state of $\aut_i$ for $i \in \{1, \ldots, p\}$,
(2)~for every $i \in \{1, \ldots, k\}$ we have $v_i = \padding$, i.e., all paths have terminated.
\end{proof}

We have shown that a QAM representing answer graphs for a given query can be efficiently constructed.
The query holds if and only if its answer graph has a path satisfying arithmetical constraints.
Now, we show that the existence of a path satisfying given arithmetical constraints can be efficiently decided on graphs represented by QAMs.
We additionally prove that we can effectively compute the minimum and the maximum over labellings of paths satisfying given arithmetical constraints.
The second result will be used in Section~\ref{s:sat-opra}.

\begin{lem}\label{l:solvingqam}
For a graph $G$ and a \qam{} $\M^Q$, let $\Pi$ be the set of paths from $S_G^{\M^Q}$ to $T_G^{\M^Q}$ satisfying $\bigwedge_{i=1}^m \lambda_i[\pi] \leq c_i$ in $G_m^{\M^Q}$.
(1)~Checking emptiness of $\Pi$ can be done non-deterministically in polynomial space in $Q$ and logarithmic space in $G$.
(2)~Computing the minimum (resp., maximum) of the value $\lambda_j[\pi]$ over all paths in $\Pi$ can be done non-deterministically in polynomial space in $Q$ and logarithmic space in $G$.
The computed extremal value is $-\infty$, $\infty$, or polynomial in $G$ and exponential in $Q$.
\end{lem}

\begin{proof}
\newcommand{\norm}[1]{\|#1\|_1}
\newcommand{\cost}{\mathbf{c}}
A \emph{vector addition system with states}  (VASS) is a $\Z^d$-labelled graph $G$, i.e., $G = (V,E)$, where $V$ is a finite set and $E$ is a finite subset of $V \times \Z^d \times V$.
Depending on the representation of labels $\Z^d$, we distinguish unary and binary VASS.
The \emph{$\Z$-reachability problem} for VASS, asks, given a VASS $G$ and its two \emph{configurations} $(s,\vec{u_1}), (t, \vec{u_2}) \in V \times Z^d$,
whether there exists a path $\pi = (s,\vec{x_0}, q_1), (q_1,\vec{x_{1}},q_2) \ldots (q_{k},\vec{x_k},t)$ in $G$ from $s$ to $t$ such that
$\vec{u_1} + \sum_{i=0}^{k} \vec{x_i} = \vec{u_2}$, i.e., $\vec{u_1}$ plus the sum of labels along $\pi$ equals $\vec{u_2}$.
In contrast to reachability in VASS, in $\Z$-reachability we allow configurations with negative components.

We discuss how to reduce the problem of non-emptiness of $\Pi$ to the {$\Z$-reachability problem} for VASS.
We transform $G_m^{\M^Q}$ into a VASS $G' = (V,E)$ over the set of nodes of $G_m^{\M^Q}$
with two additional nodes $s,t$.
We put an edge between two nodes connected node $q_1, q_2 \in V$ labelled by the label of the source node $\vec{v}$, i.e., for all $q_1, q_2 \in V$ we have $(q_1, \vec{v},q_2)$ if $\lambda_E(q_1, q_2) = 1$ and
$\vec{v} = (\lambda_1(q),\ldots, \lambda_m(q))$. Moreover,
we define $s$ as the source and $t$ and the sink, i.e.,
(1)~for every $q \in s_G^{\M^Q}$ we put an edge $(s,\vec{v}, q)$, where $\vec{v} = (c_1,\ldots, c_m)$ (constants from the definition of $\Pi$),
and (2)~for every $q \in T_G^{\M^Q}$ we put an edge $(q,\vec{v}, t)$, where
$\vec{v} = (\lambda_1(q),\ldots, \lambda_m(q))$.
Finally, we allow the labels to be increased in $t$, i.e.,
for every $i \in \{1, \ldots, m\}$, we put $(t, 1_i, t)$, where $1_i \in Z^d$ has $1$ at the component $i$, and $0$ at all other components.

Observe that paths from $\Pi$ correspond to paths in VASS $G'$ from $(s,\vec{0})$ to $(t,\vec{0})$. The $\Z$-reachability problem for unary VASS of the fixed dimension (which is $m$ in the reduction) is in \NL~\cite[Therem 19]{Blondin+15}. In the proof, it has been shown that if weights are given  in unary and there exists a path between given two configurations, then there also exists a path, which is bounded by ${p(|G'|)}^{p'(d)}$, where $p, p'$ are polynomial functions, $|G'|$ is the size of the VASS and $d$ is the dimension. This means that such a path is:
(a)~polynomially bounded in the VASS, provided that the dimension is fixed, and
(b)~bounded exponentially otherwise.

To show (1), observe that
all the labels are given in unary, and that
$m$ is fixed since $Q$ is fixed.
Therefore, if $\Pi$ is non-empty, then it contains a path of a polynomial size in $|G|$. The existence of such a path can be verified in non-deterministic
logarithmic space using the \qam{} $\M^Q$  as an oracle to query for the nodes, the edges and the labelling of $G_m^{\M^Q}$.
The \qam{} $\M^Q$ requires logarithmic space in $|G|$.
Therefore checking whether $\Pi$ is empty can be done non-deterministically in logarithmic space in $|G|$.

If $\Pi$ is non-empty, then it contains a path of the size ${p(|G'|)}^{p'(d)}$, where $|G'| = |G_m^{\M^Q}|$ is exponential in $|Q|$, but the dimension $d$ is the number of labellings in the answer graph $G_m^{\M^Q}$, and it is bounded by the number of arithmetical constraints in  $Q$.
Therefore, $d$ is  bounded by $|Q|$ and hence ${p(|G'|)}^{p'(d)}$ is exponential in $|Q|$.
Finally, if $\Pi$ is non-empty, then it contains a path of the exponential size in $|Q|$.
The existence of such a path can be verified in non-deterministic
polynomial space using  the \qam{} $\M^Q$  as an oracle to query for the nodes, the edges and the labelling of $G_m^{\M^Q}$.
Therefore checking the emptiness of $\Pi$ can be done non-deterministically in a polynomial space in $Q$.

To show (2), we need to analyse the proof of~\cite[Theorem 19]{Blondin+15}.
{The set of paths in a VASS from a configuration  $(s,\vec{v}_1)$ to $(t,\vec{v}_2)$ can be infinite,
but it can be finitely presented with \emph{path schemes}. A \emph{path scheme} is a regular expression $\rho$ over transitions of a VASS
such that every sequence of transitions matching $\rho$ is a path in the VASS.
It has been shown that there exists a finite set $S$ of path schemes of the form
$\alpha_0 \beta_1^* \ldots \beta_k^* \alpha_k$, where $\alpha_0, \ldots, \alpha_k$ are paths and $\beta_0, \ldots, \beta_k$ are cycles, such that }
(i)~each path scheme in $S$ has a polynomially bounded length in the size of VASS,
(ii)~for all configurations $(s,\vec{v}_1)$, $(t,\vec{v}_2)$, if there is a path from $(s,\vec{v}_1)$ to $(t,\vec{v}_2)$, then there is a path that matches some path scheme from $S$.
Next, it has been shown that for every  path scheme $\alpha_0 \beta_1^* \ldots \beta_k^* \alpha_k = \rho \in S$, there is a system of linear Diophantine equations $\mathcal{E}_{\rho}$ such that
$\mathcal{E}_{\rho}$ has a solution $x_1, \ldots, x_k$ if and only if
$\alpha_0 \beta_1^{x_1} \ldots \beta_k^{x_k} \alpha_k$ is a path from $(s,\vec{v}_1)$ to $(t,\vec{v}_2)$.
For each such a system of linear Diophantine equations $\mathcal{E}_{\rho}$, the set of all its solutions has a special form.
Let $cone(P_{\rho})$ be the set of linear combinations of vectors from $P$ with non-negative integer coefficients.
Then, the set of solution of $\mathcal{E}_{\rho}$ is of the form $B_{\rho} + cone(P_{\rho})$, where
$B_{\rho},P_{\rho}$ are sets of vectors whose coefficients are
(a)~exponentially bounded in the dimension,
(b)~polynomially bounded in the size of VASS (with the dimension fixed).
It follows that if $\Pi$ is non-empty, one of the following holds:
\begin{enumerate}[(1)]
\item For some path scheme $\rho = \alpha_0 \beta_1^* \ldots \beta_k^* \alpha_k$, sets $B_{\rho},P_{\rho}$ are non-empty ($\mathcal{E}_{\rho}$ has infinitely many solutions), and for some
$\vec{u} = (u_1, \ldots, u_k)$, the sum of the value $\lambda_j[\pi]$ over paths $\beta_1^{u_1}, \ldots, \beta_k^{u_k}$ is negative, and hence the minimum is $-\infty$,
\item Otherwise,
the minimum exists and it is realized by some path $\pi$ matching some path scheme $\rho = \alpha_0 \beta_1^* \ldots \beta_k^* \alpha_k$ such that
$\pi = \alpha_0 \beta_1^{x_1} \ldots \beta_k^{x_k} \alpha_k$, where $(x_1, \ldots, x_k) \in B_{\rho}$. Observe that it does not pay off to incorporate vectors from $P_{\rho}$
as they cannot decrease the value of $\lambda_j[\pi]$. Finally, observe that the size of such a path $\pi$ is polynomial in the size of VASS if the dimension of the VASS is fixed and
it is exponential in the dimension.
\end{enumerate}

\noindent
From (1) and (2), we derive bounds $b_1(G,Q) < b_2(G,Q)$, which are polynomial in $G$ and exponential in $Q$ such that
if  the minimum of $\lambda_j[\pi]$ over path in $\Pi$ exists, then it is realized by some path of length bounded by $b_1(G,Q)$.
However, if there is a path $\pi \in \Pi$ of length between $b_1(G,Q)$ and $b_2(G,Q)$, with $\lambda_j[\pi]$ lower than the value of any path shorter than
$b_1(G,Q)$,  then the minimum is infinite.
Since NL and PSPACE are closed under the complement and bounded alternation (only two conditions need to be checked), both conditions
can be checked in non-deterministically in polynomial space in $Q$ and logarithmic space in $G$.
 The case of the maximum is symmetric.
\end{proof}

\subsection{Language \opra{}}%
\label{s:sat-opra}
Assume \q{$O=\, \definedBy{\lambda_1}{t_1}$, \dots, $\definedBy{\lambda_n}{t_n}$} has depth $s$. We show by induction on $s$ that the values of the labellings of a graph $G[O]$ can be non-deterministically computed in space polynomial in $O$.

\begin{lem}\label{l:step}
Let $s$ be fixed. For a graph $G$ and $O$ of depth $s$
the value of each labelling of $G[O]$ can  be non-deterministically computed in polynomial space in $O$ and
logarithmic space in $G$.
\end{lem}

The proof studies all the possible constructors of terms.
In the case of subqueries, we apply the inductive assumption, i.e., Theorem~\ref{th:query-in-NL} for a query with fewer auxiliary labellings.
The case of the minimum follows from (2) of Lemma~\ref{l:solvingqam}, i.e., $\min_{p \in \Pi} \lambda_i[p]$ is either $-\infty$, $+\infty$ or exponential in $Q$ and polynomial in $G$.
Therefore, it can be computed using Lemma~\ref{l:solvingqam} and the bisection method. The case for the maximum is symmetric.
Finally, the application of a function symbol to terms or ranges can be implemented in the expected complexity.

\begin{proof}
The proof is by induction on $s$.
The basis of induction, $s = 0$, is trivial.
Assume that for $s$ the lemma statement and Lemma~\ref{l:induction} hold.
We show that it holds for $s+1$.
Consider a graph $G$, ontologies $O$, and
$O' =\, O$, $\, \definedBy{\lambda_1}{t_1}$, \dots, $\definedBy{\lambda_s}{t_k}$ such that $\lambda_1, \ldots, \lambda_k$ are independent, i.e.,  $t_1, \ldots, t_k$ do not refer to labellings $\lambda_1, \ldots, \lambda_k$.

We show that the value of each $t_{i}$ can  be non-deterministically computed in polynomial space in $O$ and
logarithmic space in $G$. We start the computation from the bottom, the leaves, and show that
the values of leaves can  be non-deterministically computed in polynomial space in $O$ and
logarithmic space in $G$. Indeed, leaves are of one of the following forms:
$c \mid \lambda(\vec{y}) \mid [Q(\vec{y})] \mid \min_{\lambda, \pi} Q(\vec{y}, \pi) \mid \max_{\lambda, \pi} Q(\vec{y}, \pi) \mid y = y$.
The cases of $c$ and $y = y$ are trivial.
For leaves of the form $\lambda(\vec{y})$, we know that $\lambda$ is either a graph labelling or it is from $O$ and hence we use the inductive assumption of this lemma.
Finally, for leaves of the form $[Q(\vec{y})]$, $\min_{\lambda, \pi}  Q(\vec{y}, \pi)$ and \(\max_{\lambda, \pi} Q(\vec{y}, \pi)\), it follows from Lemma~\ref{l:induction} applied inductively for $s$.

The internal nodes of $t_{i}$ are of the form
$ f(t_1(\vec{y}), \dots, t_k(\vec{y}) ) $ or $ f'(\{ t(x) \colon t(x,\vec{y})\})$.
Having the values of subterms $t_1(\vec{y}), \dots, t_k(\vec{y})$, the value $f(t_1(\vec{y}), \dots, t_k(\vec{y}) )$
can be computed in logarithmic in length of $\vec{x}$ and values in $\vec{x}$, i.e., we require space $\max(\log(|t_1(\vec{y})|), \ldots, \log(|t_k(\vec{y})|)) + \log(k) + C$, where $C$ is a constant.
Similarly, to compute  $f'(\{ t(x) \colon t(x,\vec{y})\})$, we require space $\max_x(\log(|t_1(x,\vec{y})|)) + \log(|G|) + C$, where $C$ is a constant.
It follows that to compute the value of $t_i$, we require space $|t_i| (\log|G| + C) \cdot M$, where $C$ is the maximal constant taken over all $f \in \F$ (which is fixed), and
$M$ is the maximum over space requirements of the leaves, which is logarithmic in $G$ and polynomial in $t_i$.

Since all labellings $\lambda_1, \ldots, \lambda_k$ are independent, the result follows.
\end{proof}

Finally, we sketch the proof of Lemma~\ref{l:induction}.
\begin{proof}(of Lemma~\ref{l:induction})
The proof is by induction on $s$.
 Lemma~\ref{l:pratoqam} and Lemma~\ref{l:solvingqam} imply the basis of induction.
Next, assume that this lemma holds for $s$.
Consider an  query \q{LET $O$ IN $Q'$}, with $\depth(O) = s+1$ and a graph $G$.
We first build a \qam{} $\M^{Q'}$ as in Lemma~\ref{l:pratoqam}. Since $\M^{Q'}$ may refer to labellings from $O$, not defined in $G$,
we change it so that whenever it wants to access a value of one of the labellings defined in $O$, it instead runs a procedure guaranteed by Lemma~\ref{l:step}.
Lemma~\ref{l:step} holds because this lemma holds for $s$.
Finally, we use Lemma~\ref{l:solvingqam} to determine the result.
\end{proof}

 \section{Conclusions}
We defined a new graphical query language for databases, \opra{}.
Among its advantages are good expressive power, modularity and reasonable complexity.

We demonstrated the expressive power of \opra{} in two ways.
We presented examples of natural properties and \opra{} queries expressing them in an organised, modular way.
We also showed that \opra{} strictly subsumes \ECRPQLC.
Despite the additional expressive power, the complexity of the query evaluation problem for  \opra{} matches the complexity for \ECRPQ.

\bibliographystyle{alpha}
\bibliography{bib}
\end{document}